\pdfoutput=1
\documentclass[a4paper,12pt]{article}

\usepackage{amsmath,amssymb,amsfonts}
\usepackage[dvips]{graphicx}
\usepackage{epsfig}
\usepackage{color}
\usepackage{verbatim}
\usepackage{calc}
\usepackage{bbm}
\usepackage{setspace}
\usepackage{array}
\usepackage{epstopdf}
\usepackage[caption=false]{subfig}
\usepackage{xcolor}
\usepackage[left=2.4cm,top=3.3cm,right=2.4cm,bottom=3.3cm,bindingoffset=0cm]{geometry}
\usepackage{listings}

\definecolor{dkgreen}{rgb}{0,0.4,0}
\lstdefinelanguage{Matlabex}{language=Matlab,morekeywords={do,then,ones,eigensystem,random,flip}}
\newcommand{\beq}{\begin{eqnarray}}
\newcommand{\eeq}{\end{eqnarray}}
\newcommand{\bea}{\begin{eqnarray}}
\newcommand{\eea}{\end{eqnarray}}
\newcommand{\be}{\begin{equation}}
\newcommand{\ee}{\end{equation}}

\def\de{\partial}

\def\1{\mathbbm{1}}

\def\nc{\sigma}

\def\t{\tau}
\def\nn{\nonumber  \\}
\def\tlambda{\tilde \lambda}
\numberwithin{equation}{section}

\begin{document}

\title{
\begin{flushright}\ \vskip -1.5cm {\small {IFUP-TH-2017}}\end{flushright}
\vskip 20pt
\bf{ \Large Large-$N$ ${\mathbb CP}^{N-1}$  sigma model on a finite interval and
the renormalized string energy}
}
\vskip 10pt  
\author{
Alessandro Betti$^{(1)}$,  Stefano Bolognesi$^{(2,3)}$, \\
Sven Bjarke Gudnason$^{(4)}$, Kenichi Konishi$^{(2,3)}$, Keisuke Ohashi$^{(5)}$    \\[15pt]
{\em \footnotesize
$^{(1)}$Dipartimento di Ingegneria dell'Informazione e Scienze Matematiche,}\\[-5pt]
{\em \footnotesize
 via Roma, 56, 53100 Siena, Italy}\\[3pt]
{\em \footnotesize
$^{(2)}$Department of Physics ``E. Fermi", University of Pisa}\\[-5pt]
{\em \footnotesize
Largo Pontecorvo, 3, Ed. C, 56127 Pisa, Italy}\\[3pt]
{\em \footnotesize
$^{(3)}$INFN, Sezione di Pisa,    
Largo Pontecorvo, 3, Ed. C, 56127 Pisa, Italy}\\[3pt]
{\em \footnotesize
$^{(4)}$Institute of Modern Physics, Chinese Academy of Sciences, Lanzhou 730000, China}\\[3pt]
{\em \footnotesize
$^{(5)}$ Research and Education Center for Natural Sciences,  Keio University}  \\[-5pt]
{\em \footnotesize  Hiyoshi 4-1-1, Yokohama, Kanagawa 223-8521, Japan }
 \\ [5pt] 
{ \footnotesize  stefanobolo@gmail.com, }
{ \footnotesize kenichi.konishi@unipi.it,} 
{ \footnotesize    bjarke@impcas.ac.cn,}  
{ \footnotesize   keisuke084@gmail.com}
}
\date{August 2017}

\maketitle

%\vskip 0pt

\begin{abstract}

We continue the analysis started in a recent paper of the large-$N$ 
 two-dimensional ${\mathbb CP}^{N-1}$   sigma model,   defined on a finite space interval $L$ with Dirichlet (or Neumann) boundary conditions.
Here we    focus our attention on   the problem of the renormalized energy density  $ {\cal E}(x, \Lambda, L)$ which  is found to be a sum of two terms, 
a constant term coming from the sum over modes, and a term proportional to the mass gap.  The approach to $ {\cal E}(x, \Lambda, L)   \to       \tfrac{N}{ 4  \pi }  \Lambda^2 $   at large $L \Lambda$ is shown, both analytically and numerically,
 to be exponential: no power corrections are present and in particular no L\"uscher term appears.
This is  consistent with the earlier result which states that the system
has a unique massive phase, which interpolates  smoothly between the classical weakly-coupled limit  for  $L \Lambda \to 0$ and the  ``confined" phase of the standard  ${\mathbb CP}^{N-1}$ model in two dimensions for $L \Lambda \to \infty$.
%No spontaneous breakdown of the isometry group $SU(N)$ takes place. 

\end{abstract}

\newpage

\section{Introduction}

   Recently we embarked on the investigation of the bosonic ${\mathbb CP}^{N-1}$ model \cite{D'Adda:1978uc,Witten:1978bc}, defined on finite space interval $L$,  i.e.,   on a finite width worldstrip,  in the large $N$ approximation \cite{BKO}. Such a system could provide a useful model for various physical situations.  For instance, 
it appears as the low-energy effective theory describing the quantum
excitations of the monopole-vortex soliton complex  \cite{MVComplex1,MVComplex2,MVComplex3,MVComplex4} in hierarchically broken gauge symmetries, such as  $SU(N+1) \to SU(N) \times U(1) \to {\bf 1}$ in a color-flavor locked $SU(N)$ symmetric vacuum.   The ${\mathbb CP}^{N-1}$  model  describes the nonAbelian orientational zeromodes of the nonAbelian vortex (string) \cite{Hanany:2003hp,Auzzi:2003fs,Shifman:2004dr},  whereas its boundaries represent the monopoles arising from a higher-scale gauge-symmetry breaking, carrying the same orientational ${\mathbb CP}^{N-1}$ 
moduli. NonAbelian monopoles, not plagued by the well-known difficulties, could  emerge in such a context.  
The fate of the nonAbelian monopoles as a quantum mechanical entity is then linked to the phase of the low-energy  ${\mathbb CP}^{N-1} $  effective action attached to it.

In \cite{BKO}  it was found that the quantum saddle-point equations
describing the ${\mathbb CP}^{N-1}$ model with Dirichlet or Neumann
boundary conditions has a unique solution under certain conditions.   In the large-$L$ limit,
this solution approaches smoothly the well-known confining phase of
the standard 2D ${\mathbb CP}^{N-1}$ system.  A phase transition
between a Higgs-like phase and the confining phase for a shorter $L$, which was claimed   to be present in the literature \cite{Milekhin:2012ca}, 
was shown  not to exist in the system.\footnote{For periodic boundary
  conditions (and large-$N$), however, such a phase transition 
  does occur \cite{Monin:2015xwa}.  See also  \cite{Milekhin:2016fai}. }

The model is interesting also from a formal point of view, as it
provides a prototype model of a quantum system of varying 
dimensions in the presence of dynamical mass generation: it interpolates between  a  2D QFT  (in the $L\to \infty$ limit)  with all well-known phenomena such as asymptotic freedom and 
confinement  and a 1D system in the $L \to 0$ limit - quantum mechanics. 
For shorter strings of length $L \le  1/\Lambda$, quantum fluctuations
of the ${\mathbb CP}^{N-1}$ fields $n_i$,  ($i=1,2, \ldots,  N$)
remain weakly coupled,  as they lack sufficient 2D spacetime
``volume'' in which the fields fluctuate. With the Dirichlet
condition, the system reduces effectively to a classical system in the
$L \ll 1/\Lambda$ limit.

In this paper, we delve in more detail into the properties of the large-$N$ ${\mathbb CP}^{N-1}$ model on a finite-width worldsheet. 
First, with a more refined numerical method we improve the precision
of the solution to the generalized gap equation. This enables us to explore a larger region of the parameter space and in particular the limit of large $L$.
The second problem is to understand
 the energy density of the string itself as a function of $x$, computed at the functional saddle point,  completing the analysis presented  in  \cite{BKO}.  
The third problem is to 
 clarify the approach to the QFT ($L \to \infty$) limit of our system;
 this involves the question of certain consistency with the known
 field-theory limit, as well as of figuring out interesting $L$-dependent effects. It will be seen that power-behaved corrections such as the L\"uscher term are absent.  This is consistent  as all fields acquire dynamically generated mass; at the same time no spontaneous breakdown of the global $SU(N)$ symmetry takes place. 
 
The paper is organized as follows. 
In Section \ref{two} we review the ${\mathbb CP}^{N-1}$  model on a finite strip and also present new numerically improved results which allows to reach higher values of $L$ than before.
In Section \ref{sec:gapeq}  the generalized gap equation is re-derived,   paying special attention to the anomalous term that arises in the functional variation, which is analogous to the axial anomaly.
In Section \ref{four} we consider the energy density in detail, its various contributions, and its  $L \to \infty$ limit.
In Section \ref{sec:Analytic} we study an analytical Ansatz that describes the large but finite  $L \gg 1/\Lambda$ and the approach to the $L\to \infty$ limit. 
The numerical results for the renormalized energy density are presented in Section \ref{six}. 
In Section \ref{seven} we discuss the Casimir force.  Our conclusion is in
Section \ref{eight}.   Some details of our analysis are given in Appendices \ref{app:prop} $\sim$ \ref{app:algorithm_test}.

\section{Review of the ${\mathbb CP}^{N-1}$  model on a finite width worldsheet}
\label{two}

The classical  action for the ${\mathbb CP}^{N-1}$ sigma model is defined  by 
\beq
S= \int dx dt \left( (D_{\mu} n_i)^*D^{\mu} n_i -  \lambda (n_{i}^* n_i - r)   \right) \;,\qquad  r= \frac{4 \pi}{g^2}\;, 
\eeq
where $n_i$ with $i=1,\dots,N$ are $N$ complex scalar fields and the
covariant derivative is given by $D_\mu=\partial_\mu-i g A_\mu$.
Configurations related by $U(1)$ gauge transformations
$n_i \to e^{i\alpha} n_i$ are not only gauge-equivalent, but
  are equivalent because the $U(1)$ gauge field
$A_{\mu} $ does not have a kinetic term in the classical action. 
$\lambda$ is a Lagrange multiplier field that enforces the classical
gap equation
\beq
\label{classicalconstrain}
n_{i}^* n_i  = r\;,     \eeq
where $r= \tfrac{4\pi}{g^2}$ is related to the gauge coupling and can
be thought of as the ``size'' of the ${\mathbb CP}^{N-1}$ manifold.

For the ${\mathbb CP}^{N-1}$ theory on a finite interval of length $L$,  $x \in[-\tfrac{L}{2},\tfrac{L}{2}]$,\footnote{With the aim of studying the $L \to \infty $ 
limit of the string at fixed  $x$ (and $\Lambda$)  in mind,  we take the space interval to be  $[-\tfrac{L}{2},\tfrac{L}{2}]$, rather than $[0,   L]$   as done  in \cite{BKO},  by a trivial shift of the 
spatial coordinate.}
    the boundary conditions must be specified. One possibility is the
    Dirichlet-Dirichlet boundary condition which -- up to a $U(N)$
    transformation -- is 
\beq
\hbox{D-D}: \qquad
n_1\!\left(-\tfrac{L}{2}\right)=n_1\!\left(\tfrac{L}{2}\right) =
\sqrt{r}\;,   \qquad
n_{i}\!\left(-\tfrac{L}{2}\right)=  n_{i}\!\left(\tfrac{L}{2}\right) =0\;,  \quad i>1\;.  \label{DDbc}
\eeq
For the moment we take the boundary conditions for the $n_i$ fields in the same direction in the ${\mathbb CP}^{N-1}$ space at the two boundaries.  
Another possibility is the Neumann-Neumann  boundary condition\footnote{
Mixed conditions can be chosen  
where one of the boundaries takes the Dirichlet condition and the other the Neumann condition.  
}
\beq
\hbox{N-N}: \qquad \partial_x n_i\!\left(-\tfrac{L}{2}\right) =
\partial_x n_i\!\left(\tfrac{L}{2}\right) =0\;, \quad \forall i. \label{NNbc}
\eeq
In this paper, we will focus on the Dirichlet-Dirichlet boundary
condition. Thus with this condition the $N$ fields can naturally be
separated into a classical component $\nc \equiv n_1$ and the rest,
$n_i$    ($i=2,\dots,N$). 
Integrating out the $n_i$ fields yields the effective action:
\beq  
{S}_{{\rm eff}} = \int d^2 x \left( (N-1) \, {\rm tr}\, {\rm log} (- D_{\mu}D^{\mu}  + \lambda) + (D_{\mu} \nc)^*D^{\mu} \nc -  \lambda (|\nc|^2  - r)  \right)\;.\label{from}
\eeq
Because the effective action only depends on $|\sigma|^2$ and
$|\partial_\mu\sigma|^2$ and the boundary conditions take real
positive values on both sides, one can take $\sigma$ to be a real
field and set the gauge field to zero. 
Finally, we will consider the leading contribution at large $N$ only. 

The generalized gap equations following from Eq.~(\ref{from})  (see also Section~\ref{sec:gapeq} below)
\beq
\frac{N}{2} \, \sum_n\frac{f_n(x)^2}{\omega_n} e^{-\epsilon \omega_n}  + \nc(x)^2 - r_{\epsilon} = 0  \,,\qquad  \partial_x^2 \nc(x) - \lambda(x) \nc(x) = 0  \,,  
\label{gapeqbb}
\eea
where
\begin{eqnarray}
r_\epsilon \equiv 
\frac{N}{2\pi}\left(\log\left(\frac{2}{\Lambda  \epsilon}\right)
-\gamma \right)\;,  
 \label{uvbeta}
\end{eqnarray}
have been solved numerically  in \cite{BKO}, by a Hartree-like self-consistent method. 
The renormalized, finite functions of $x, \Lambda, L$, $\lambda(x)$ and $\sigma(x)$ 
have been obtained numerically for various  values of $L $ and $\Lambda$. 
The calculations of  \cite{BKO} have  been extended to larger values
of $L$, with a considerably improved method.
The  weak point of the
Hartree-like method  -- from a numerical point of view --  is the need  to determine $\lambda(x)$ from the second
equation in Eq.~\eqref{gapeqbb}, where  $\sigma$ tends to zero in the
middle of the string.

It is sometimes convenient to rewrite the first equation in Eq.~(\ref{gapeqbb}) as  
\beq
N D(x,\epsilon; x,0)  + \nc(x)^2 - r_{\epsilon} = 0 \;,  \label{Dgapeq}
\eeq
  in terms of the two-point function
\begin{eqnarray}
 D(x,\tau; x',\tau')\equiv \sum_n\frac{e^{-|\tau-\tau'| \omega_n}}{2\omega_n} f_n(x) f_n(x')\;.   \label{thetwo}
\end{eqnarray}
The latter satisfies,  for the D-D boundary condition, an equation  
\beq
&& \left(-\partial_\tau^2 -\partial_x^2+\lambda(x) \right) D(x,\tau;x',\tau')\nn
&=& \sum_{n\in \mathbb Z} 
\delta(\tau-\tau')\left\{ \delta(x-x'+2n L)-\delta(x+x'+(2n+1) L) \right\}. \label{Dequation}
\eeq
Note that the infinite number of mirror poles are required to satisfy the D-D boundary condition. 
See Appendix \ref{derivation} for more details.

Under the assumption 
\beq
\lim_{x\to \pm \frac{L}2} \left(x\pm \frac{L}2\right)^2 \lambda(x) =0\;, \label{asymptotic-free}
\eeq 
 the near-the-boundary behavior of the fields turns out to be \cite{BKO} (see Section~\ref{section:near} below):   
\beq
 \sigma^2 \simeq   \frac{N}{2 \pi}  \log{\frac{1}{|x\pm L/2|} }  \;; 
\qquad  
\lambda(x) \simeq    \frac{1}{2  \,(x\pm L/2)^2 \log{1/|x\pm L/2|} } \;. \label{satisfying}
\eeq

\subsection{Numerical method and solutions}

The new method is based on a random-walk algorithm and is reversed in
some sense with respect to the old method.
A guess can be made for the function $\lambda(x)$, but the precise
starting point is not important.
The algorithm has two assumptions built in; basically just for saving
computational costs; i.e.~$\lambda(x)$ is a symmetric function in 
$x$; the second is that $\lambda(x)$ is a monotonically increasing
function from the midpoint of the string to the boundary (both
assumptions are indeed consistent with the results of \cite{BKO}). 
Now the algorithm makes a random change to a part of the function
$\lambda(x)$ (viz.~on an interval that is a randomly chosen subset of
the full string interval) yielding $\tilde{\lambda}(x)$.
Now the new $\tilde{\lambda}(x)$ function is tested in the following
way. $\sigma(x)$ is calculated from its equation of motion (second
equation of Eq.~\eqref{gapeqbb}) with the appropriate Dirichlet
boundary conditions as well as from the generalized gap 
equation (the first equation of Eq.~\eqref{gapeqbb}); these two are
compared.
If the new $\tilde{\lambda}(x)$ makes the two $\sigma$s move closer to
each other, then the new $\tilde{\lambda}(x)$ function is accepted as
the new improved $\lambda(x)$, otherwise it is rejected.
Then the cycle repeats until the precision is good enough (for the
solutions we found, that is
$\int_{-L/2}^{L/2} dx\;|\sigma^2-\tilde{\sigma}^2|<10^{-5}$).
See Appendix~\ref{sec:walk} and Appendix~\ref{app:algorithm_test}  for more details.

\begin{figure}[!ht]
\begin{center}
\mbox{\subfloat{\includegraphics[width=0.49\linewidth]{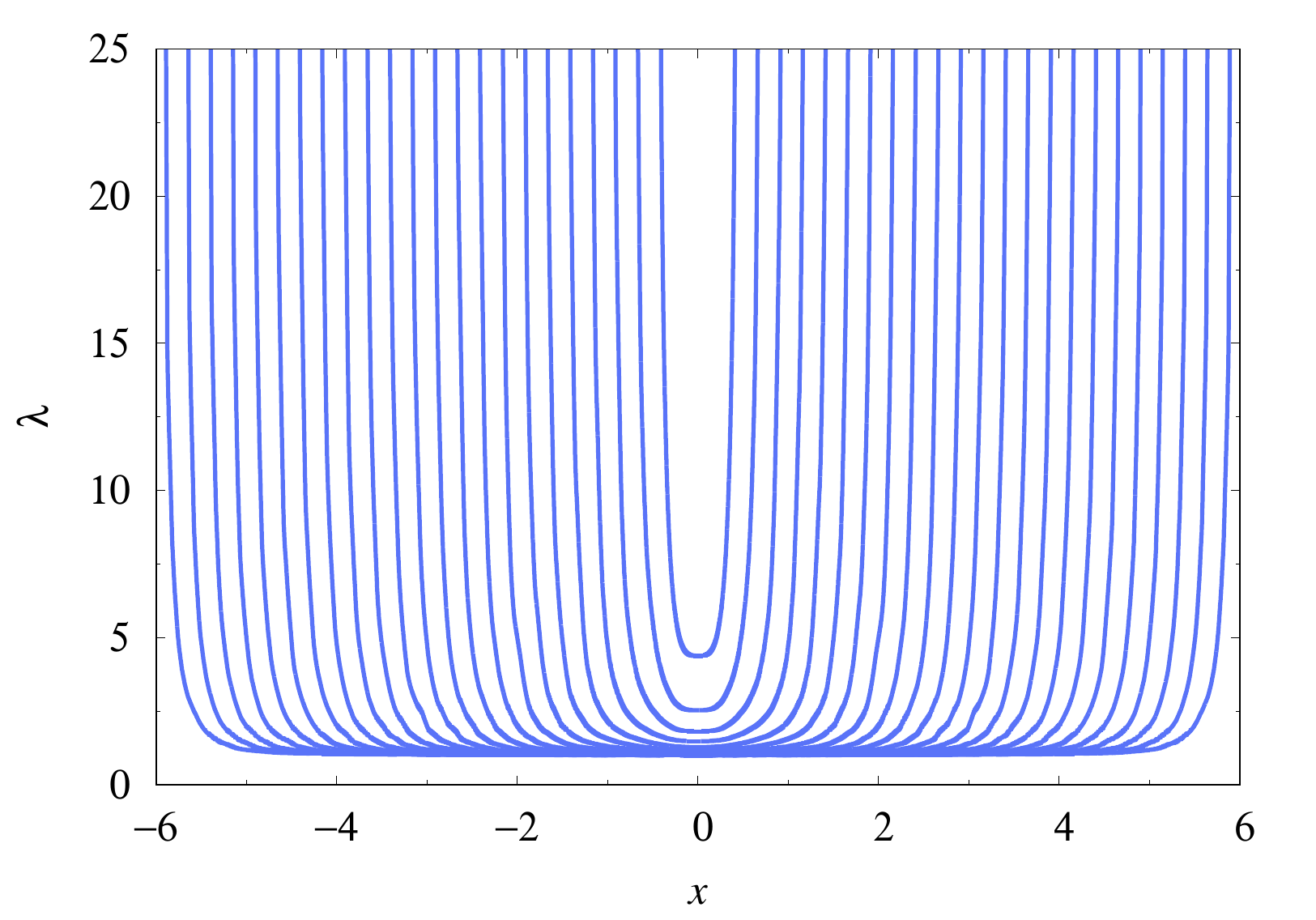}}
\subfloat{\includegraphics[width=0.49\linewidth]{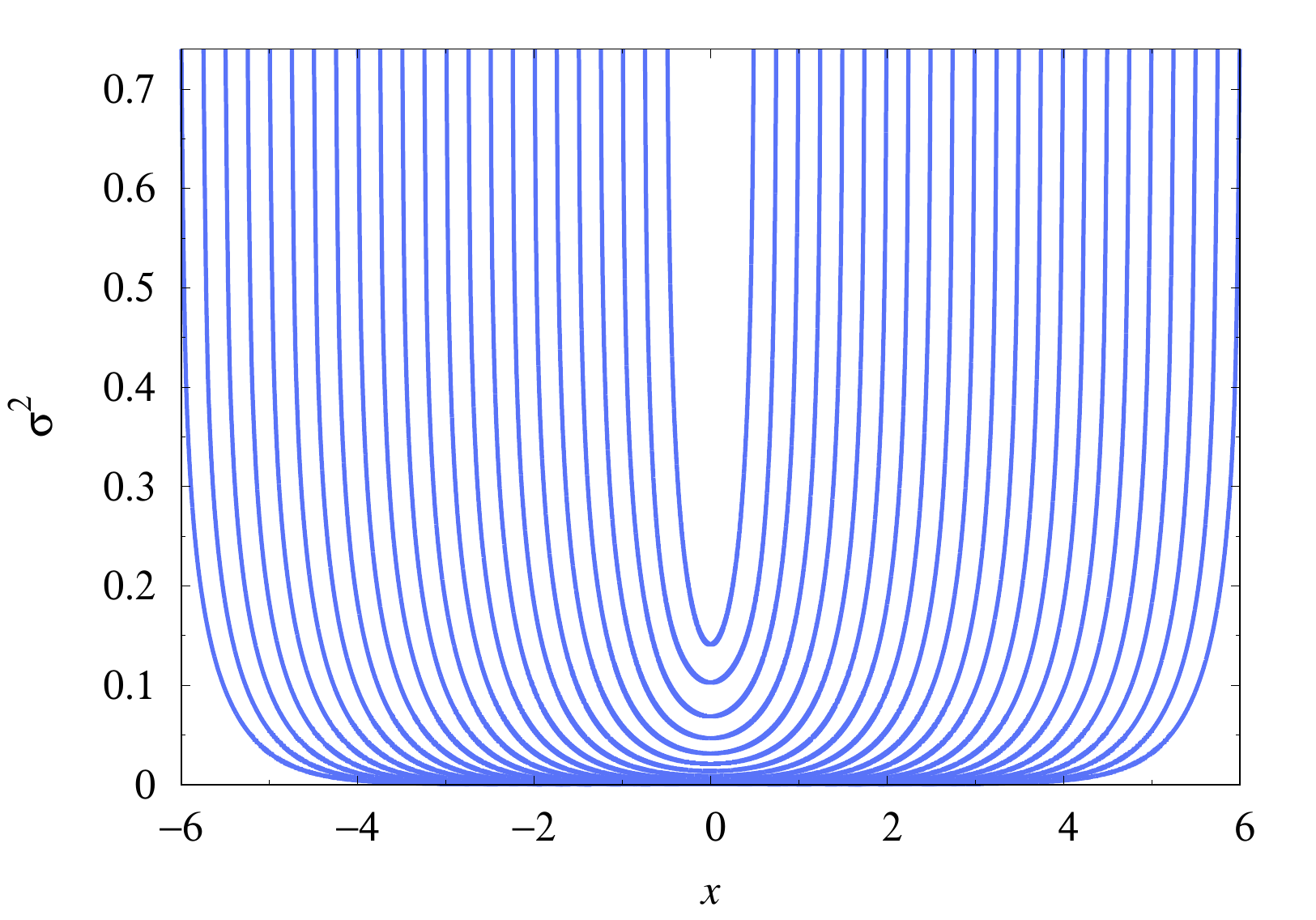}}}\vspace{1em}\\
\caption{The functions  $\lambda(x)$ (left) and   $\sigma^2(x)$ (right)  which are solutions to the gap equation, Eq.~(\ref{gapeqbb}), for 
various values of $L$  ranging  $L=1 \sim 12$.   $\Lambda=1$ in this figure.
The innermost (outermost) curve corresponds to $L=1$ ($L=12$).
}
\label{fig1}
\end{center}
\end{figure}

\begin{figure}[!ht]
\begin{center}
\mbox{\subfloat{\includegraphics[width=0.49\linewidth]{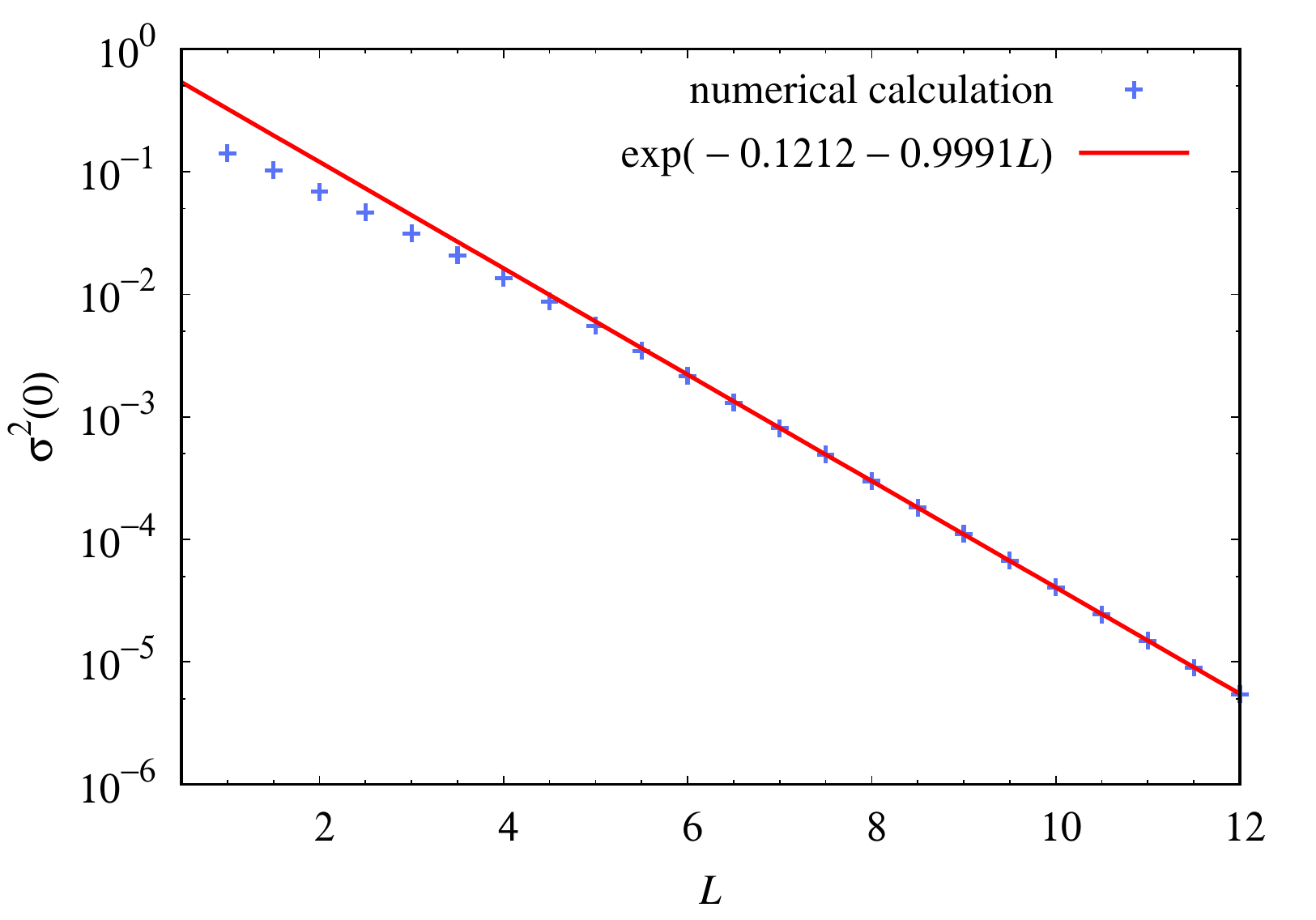}}
\subfloat{\includegraphics[width=0.49\linewidth]{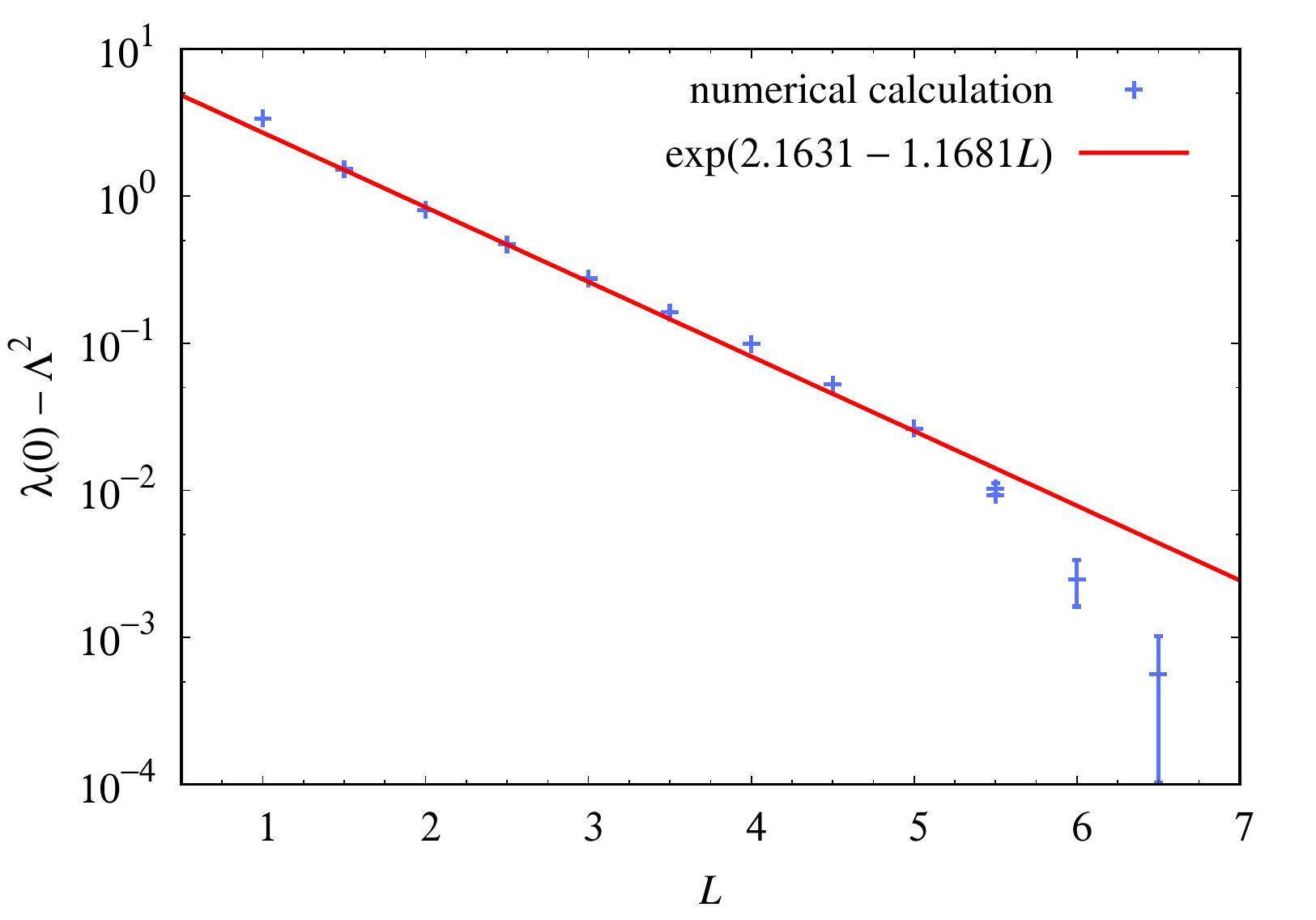}}}\vspace{1em}\\
\caption{The values of fields in the middle of the interval
  $\sigma^2(0)$ and $\lambda^2(0)-\Lambda^2$ in a logarithmic plot. $\Lambda=1$ in this figure.
  These figures are just an illustration of the exponential behavior of $\sigma^2(0)$ and $\lambda^2(0)-\Lambda^2$ as functions of $L$, and are not intended 
  to be precise fits to theory. Especially the numerical errors are quite large at large $L$ in the right figure (but still much smaller than $\Lambda^2=1$).
}
\label{fig2}
\end{center}
\end{figure}

\begin{figure}[!ht]
\begin{center}
\mbox{\subfloat{\includegraphics[width=0.49\linewidth]{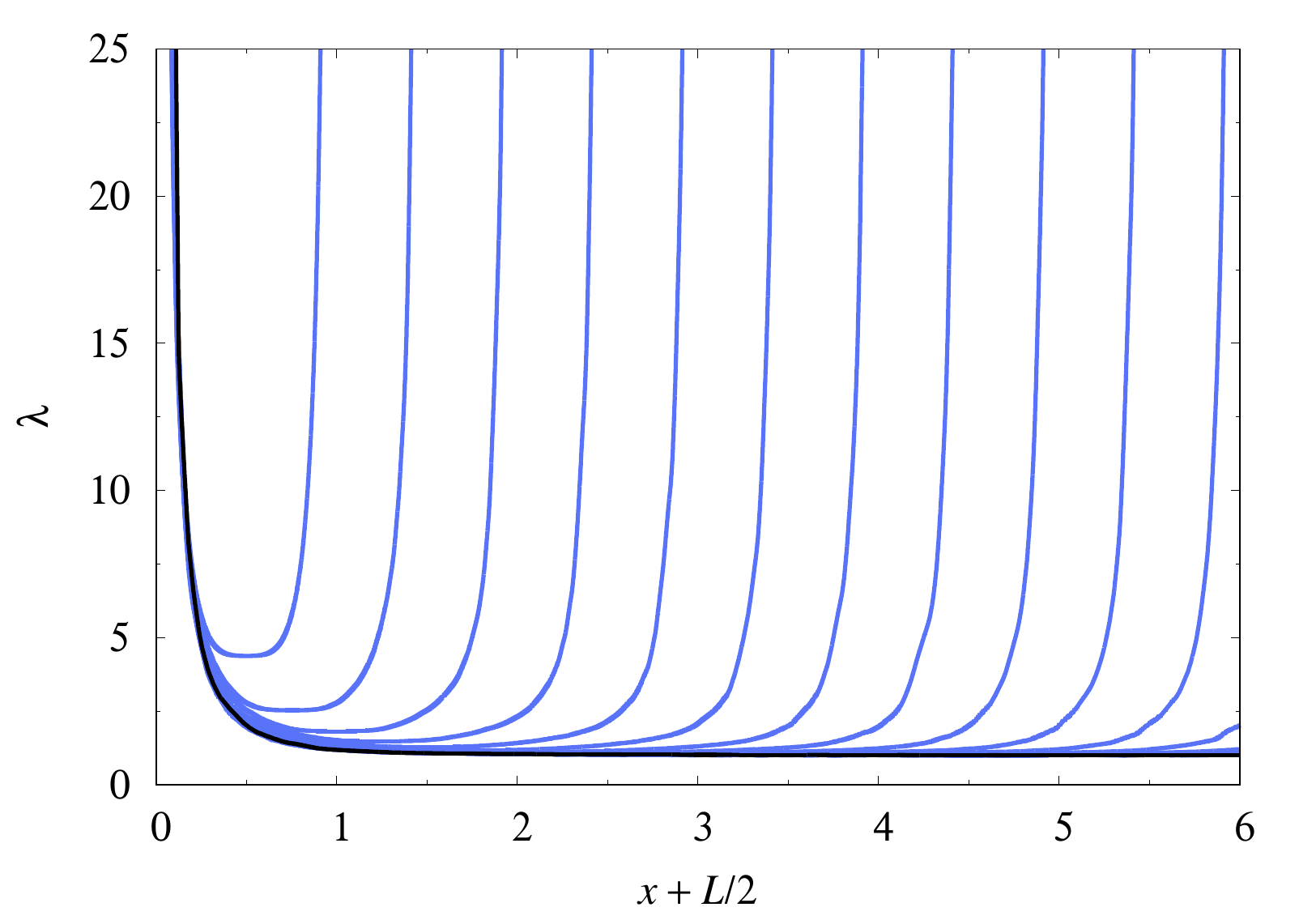}}
\subfloat{\includegraphics[width=0.49\linewidth]{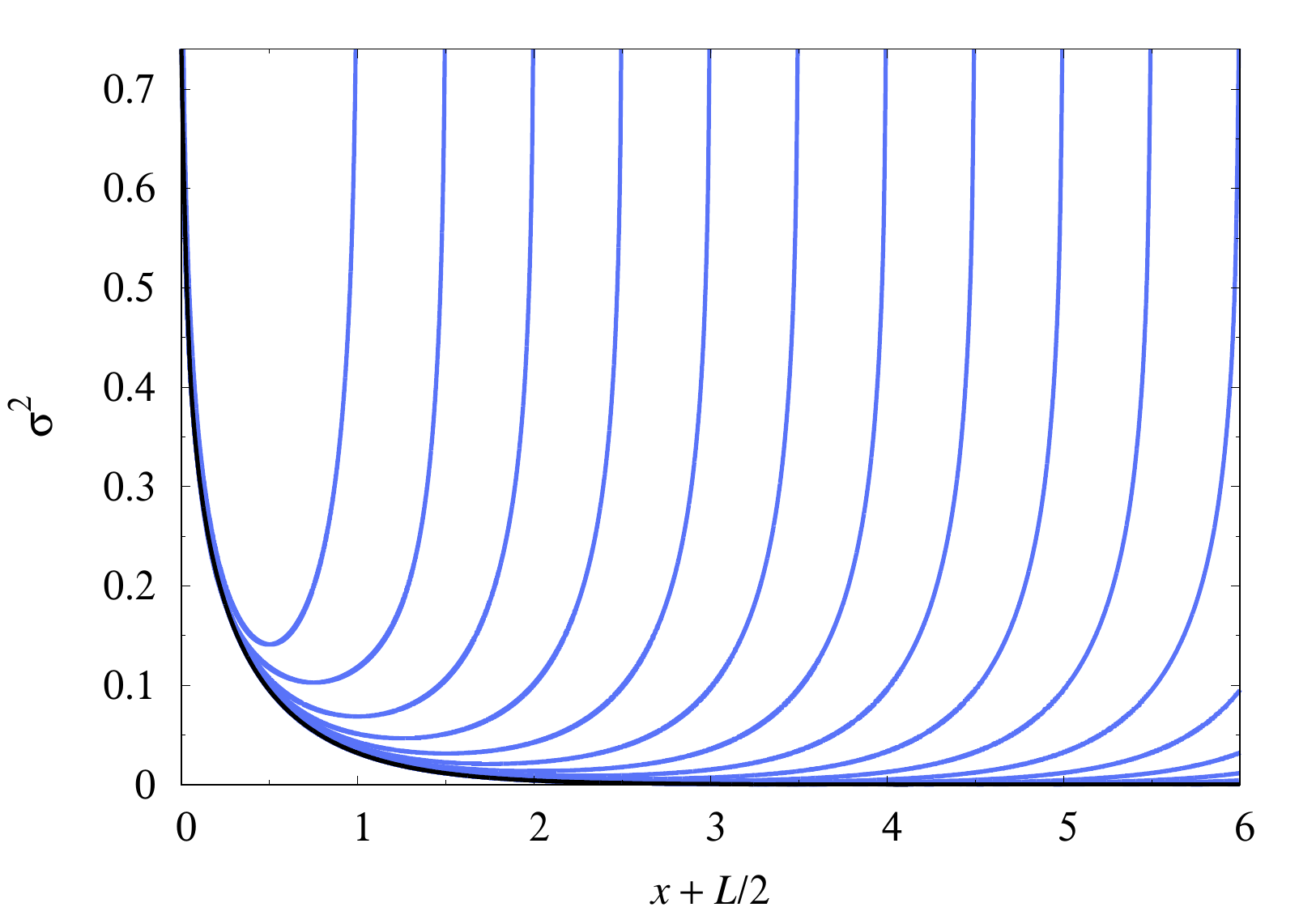}}} \vspace{1em} \\
\caption{The function  $\lambda(x)$ (left) and   $\sigma^2(x)$ (right)  which are the solutions of the gap equation, Eq.~(\ref{gapeqbb}) in the interval $(-\frac{L}{2},0]$ for 
various values of $L$  ranging  $L=1 \sim 12$. $\Lambda=1$ in this figure.
}
\label{fig3}
\end{center}
\end{figure}

Some  examples  are shown in Figure \ref{fig1} where the approach to the confined phase at large $L$ is evident.
The values of fields in the middle of the interval $\sigma^2(0)$ and $\lambda^2(0)-\Lambda^2$ are shown in a logarithmic plot in Figure \ref{fig2}.
The approach of $\sigma^2(0)$ to zero is clearly exponential and consistent with its mass.
In Figure \ref{fig3} we show the solutions in the interval $(-\frac{L}{2},0)$ by keeping one boundary fixed at $-\frac{L}{2}$.  This clearly shows the convergence at large $L$ to the half-line solution $(-\infty,0]$.

As already noted in \cite{BKO},  one can clearly see  that the
asymptotic ($L\to \infty$) regime  (with $\lambda(x)\sim \Lambda^2$
and $\sigma(x)\sim 0$   except near the boundaries)  has already set
in  at $L\simeq  4$, which is quite reasonable ($L \Lambda=4 \gg 1$).
The effect of the boundaries is seen to propagate only for $\Delta x
\sim  1/\Lambda$ from the latter:   the system effectively reduces to
the standard  2D ${\mathbb CP}^{N-1}$   model in an infinite spacetime,  as one moves away from the boundaries by $\sim1/\Lambda$ or more,   as expected on general grounds.

 \subsection{$L$ dependence, near-the-boundary  behavior of $\lambda(x), \sigma(x)$, and the classical limit   \label{section:near}}

Our system has two parameters, the dynamically generated mass scale $\Lambda$ and the interval length $L$.  But as $\Lambda$ fixes the physical unit of length, the model actually possesses only one parameter: what distinguishes two physically distinct systems is the product $\Lambda L$.
The crucial point is that the UV divergences 
are  short-distance effects around any fixed space point, and are universal. They do not depend on  the presence or absence of the boundaries. 
This is what allows us to define unambiguously systems of ``different space width $L$".

The near-the-boundary behavior (\ref{satisfying}) can be obtained  \cite{BKO} as follows.  Under the assumption (\ref{asymptotic-free}), 
the large $n$ modes are given at any fixed $x$  by 
 \beq
\label{largen}
f_n \simeq 
\sqrt{\frac2L} \sin\left(\frac{n \pi (x+ L/2)}L\right)\;,  \qquad  
\omega_n \simeq \frac{n\pi}L\;, \qquad n \gg 1\;.
\eeq
The finite part of the sum over modes in the gap equation behaves then as
\begin{equation}
\frac{N}{2} \, \sum_n\frac{1}{\omega_n}\left( f_n(x)^2 -\frac1L \right)  =
 \frac{N}{2 \pi} \log
\left(2 \sin{\left(\frac{\pi (x+L/2)}{L} \right)}  \right)\simeq     \frac{N}{2 \pi}   \log \left(\frac{2\pi}{L}(x+L/2)\right)\; ,  \label{massless}
\end{equation}
near the left boundary.  This singularity can only be compensated by $\sigma(x)^2$ in the gap equation, hence  Eq.~(\ref{satisfying}).  The numerical solutions found in \cite{BKO} and here clearly exhibit this logarithmic behavior.

There is an alternative way of understanding  the behavior of $\sigma(x)$ and $\lambda(x)$ near the boundaries. Consider the regularized but un-renormalized form of the gap equation (\ref{gapeqbb}).  By keeping the UV regularization parameter $\epsilon $ fixed and by going to $x=\mp  \tfrac{L}{2}$, one finds   that 
\beq   \sigma\!\left(\mp \tfrac{L}{2}\right)^2 =    r_{\epsilon} =  \frac{N}{2\pi}   \log\left(\frac{1}{ \epsilon} \right) +\cdots \;,  \label{this}
\eeq
as $f_n=0$.   This is nothing but the classical ${\mathbb CP}^{N-1}$ model (\ref{classicalconstrain}) with the Dirichlet boundary condition
    \beq
   n_1\!\left(-\tfrac{L}{2}\right)=n_1\!\left(\tfrac{L}{2}\right) = \sqrt{r}\;,   \qquad   n_{i}\!\left(-\tfrac{L}{2}\right)=  n_{i}\!\left(\tfrac{L}{2}\right) =0\;,  \quad i>1\;. 
\eeq
With $x$  close to but not exactly  at  a boundary, $\epsilon$  in (\ref{this})  is replaced by 
 $\left|x\pm\tfrac{L}{2}\right|$ and one finds Eq.~(\ref{satisfying}).  This statement requires an explanation. 
 What really happens is that   in the gap equation (\ref{gapeqbb}),      $\log 1/\epsilon$  which is in  $\sigma(x)$ at  exactly $x=-L/2$,    is moved  at small but nonvanishing  $|x+L/2|$  to  the first term involving the sum over the modes.  After $\log 1/\epsilon$
 is eliminated by the bare coupling constant term
 $r_{\epsilon}$ and the gap equation is renormalized and made finite, 
    it produces $- \log 1/ |x+L/2| $,    which can only be compensated by
    $\sigma(x)^2$, as  in   (\ref{satisfying}).  See
    Eq.~(\ref{both3})  and Eq.~(\ref{both4}).

This discussion clearly shows that the origin of the singular behavior of the mass gap  $\lambda(x)$  and of the $\sigma(x)$ field is the fact that  the system reduces to its classical limit \footnote{We thank Misha Shifman for useful discussions on this point.} near the boundaries, not having sufficient 2D spacetime volume for the $n_i$ fields to fluctuate. 

The same reasoning  explains \footnote{These two issues are indeed one
  and the same:  in the small-$L$ limit, the system consists of its boundaries only, so to speak.}   the behavior of  the value of $\lambda(x)$ and $\sigma(x)$ at the midpoint of the string  at small  $L \ll 1/\Lambda$,  found in  \cite{BKO} (see Fig. 3 there), 
\be  \lambda(0)   \sim   \frac{4}{L^2}  \log\frac{1}{L}\;, \qquad      \sigma^2(0) \sim   \frac{N}{2\pi}\log\frac{1}{L}\;. \label{decrease}
\ee

\section{Anomalous functional variation and the generalized gap equation  \label{sec:gapeq}}

We now re-derive the generalized gap equation.
Our starting point is  the  energy density  
\begin{eqnarray}
 {\cal E}(x)&=&\frac{N}2 
\sum_{n}\left(\omega_n f_n(x)^2+\frac{1}{\omega_n}\left(f'_n(x)^2+\lambda
	 f_n(x)^2\right)\right)e^{-\epsilon \omega_n} \nonumber   \\ 
&&\mathop+\sigma'(x)^2+\lambda(x)\left(\sigma(x)^2-r_\epsilon^0\right)+{\cal
           E}_{\rm uv} \; ,  \label{density}
\end{eqnarray}
where $\epsilon$ has been introduced as a regulator of the UV divergences coming from higher modes,  
${\cal E}_{\rm uv}$ is a subtraction constant and $f_n(x)$, $\omega_n$
are the eigenmodes of the $n_i$ field equations  
\be 
 -f''_n(x)+\lambda(x)f_n(x)=\omega_n^2 f_n(x)\;,\qquad 
\int_{-L/2}^{L/2} dx \; f_n(x)f_m(x)=\delta_{n,m} \;.    \label{using}
\ee
By integrating Eq.~\eqref{density} over
$x \in\left[-\tfrac{L}{2},\tfrac{L}{2}\right]$  and by using  (\ref{using}),  one has the expression for the integrated energy, 
\begin{eqnarray}
 E   &\equiv & \int_{-L/2}^{L/2} d x\;  {\cal E}(x)\nn
&=&  N \sum_{n}\omega_n e^{-\epsilon \omega_n}
+\int_{-L/2}^{L/2}  d x \left[\sigma'(x)^2+\lambda(x)(\sigma(x)^2-r_\epsilon^0 )+{\cal E}_{\rm uv}   \label{total}
\right]\;.
\end{eqnarray} 
For instance the total derivative terms vanish as  \cite{BKO}
\begin{eqnarray}
  \lim_{x\to - L/2} f_n(x)f_n'(x)
  = \lim_{x\to -  L/2} \left(\frac{x+L/2}{-\log(x+L/2)}
  + {\cal O}\left(\frac{x+L/2}{2\log^2(x+L/2)}\right) \right)=0\;, \quad \forall n\;.
\end{eqnarray}
By varying (\ref{total})  with respect to  $\lambda(x)$, and by using 
\beq
\delta \omega_n^2 = \int_{-L/2}^{L/2}  dx \;  \delta \lambda(x)  f_n(x)^2 \,,\qquad    \frac{\delta \omega_n } {\delta \lambda(x) }=  \frac{f_n(x)^2}{2 \omega_n}\,,  
\eeq
  one gets  the generalized gap equation
\beq
\frac{N}{2} \, \sum_n\frac{f_n(x)^2}{\omega_n} e^{-\epsilon \omega_n}  + \nc(x)^2 - r_{\epsilon} = 0  \,,
\label{gapeq}
\eea
where
\be      r_{\epsilon}  = r_{\epsilon}^0 
 +\frac{N}{2\pi}  \;,   \label{anomaly}
\ee
whereas the variation with respect to $\sigma$  gives
\beq
\partial_x^2 \nc(x) - \lambda(x) \nc(x) = 0  \ . \label{lambda}
\eeq

Note the extra $\frac{N}{2\pi} $  term  in (\ref{anomaly}).  It  arises when the variation $\delta/\delta \lambda(x)$  acts on the regulator  factor $e^{-\epsilon \omega_n}$:
\be   N \sum_{n}\omega_n  \frac{\delta }{\delta \lambda(x)}  e^{-\epsilon \omega_n} =  
 -  \frac{\epsilon    N}{2}   \sum_{n}  f_n(x)^2 e^{-\epsilon \omega_n} :
\label{anvar}\ee
this term is superficially of the order of  $\mathcal{O}(\epsilon)$:   however the sum in the last expression diverges as $\frac{1}{\epsilon}$, so it gives 
a nonvanishing 
contribution \footnote{This is  exactly as the axial anomaly arises when the spacetime derivatives act on the string bit in the point-split axial current operator.}.
 As the divergence comes from large $n$ it may be 
calculated as
 \be -  \frac{\epsilon    N}{2}   \sum_{n}  f_n^2(x)  e^{-\epsilon \omega_n} \simeq -  \frac {\epsilon  N}{ 2 L}   \sum_{n} \left[  1 -  \cos \left(\frac{ 2  n \pi (x+L/2)  }L\right) \right]   e^{-\epsilon \pi n /L} \;=  - \frac{N}{2\pi} +  \mathcal{O}(\epsilon)\;,    \label{both1}   
\ee
for
 \be   - \tfrac{L}{2} < x < \tfrac{L}{2}\;,     \ee  
where use was made of an approximate form for the
eigenmodes %wave functions
 \beq
\label{largenBis}
f_n \simeq 
\sqrt{\frac2L} \sin\left(\frac{n \pi (x+ L/2)}L\right)\;,  \qquad  
\omega_n \simeq \frac{n\pi}L\;, \qquad n \gg 1\;.
\eeq
The same result  can be found
by using the propagator representation  (\ref{thetwo}).
See Appendix \ref{derivation}.

%\subsection{Anomalous term at the boundaries} 
%
%Note that either from (\ref{both1}) or from  (\ref{both3}), one sees that the anomalous term  is  absent for the gap equation at  exactly $x=0$ or $x=L$.  At small $x$, one finds that the anomalous piece in the gap equation is 
%\be     \lambda(x)   \frac{N}{2\pi}  \left( -1 +  \frac{\epsilon^2}{4 x^2 + \epsilon^2} \right)  
%\to   \begin{cases}
%       - \frac{N}{2\pi}  \lambda (x)  &        \epsilon \to 0\;\,\,,   \,\, x \ne 0   \\
%   0     &   x \to 0  \;\,\,, \,\, \epsilon \ne 0\;. 
%\end{cases}\;.
%\ee
%This does not solve the problem of divergence of the energy  density at $x\sim 0, L$.  
%The behavior of $\lambda(x)$ 
%\beq
%\lambda(x) \simeq    \frac{1}{2  \,x^2 \log{1/x}}  \; \label{satisfying}
%\eeq

%%%%%%%%%%%%%%%%%%%%%%
%%%%%%%%%%%%%%%%%%%%%%
%%%%%%%%%%%%%%%%%%%%%%
%%%%%%%%%%%%%%%%%%%%%%

\section{Energy  density \label{four}}

We now go back to the density itself, and rewrite (\ref{density})  as
\bea
    {\cal E}(x)  &=&    {\cal E}_0(x)
    +  \lambda(x)  \left( \frac{N}{2}  \sum_{n} \frac{1}{\omega_n}   f_n(x)^2       +     \sigma(x)^2-r_\epsilon^0   \right)   +{\cal E}_{\rm uv}\;  \nonumber \\
 &=&     {\cal E}_0(x)   + \frac{N}{2\pi} \lambda(x) +  {\cal E}_{\rm uv}\;,   \label{Edensity}
\eea
where 
\be   {\cal E}_0(x)  \equiv  \frac{N}{2}  \sum_{n}\left(\omega_n f_n(x)^2+\frac{1}{\omega_n}f'_n(x)^2  
	 \right)e^{-\epsilon \omega_n}   +\sigma'(x)^2  \;,    \label{EdensityConst}
\ee
by collecting terms proportional to $\lambda(x)$ and by using Eqs.~(\ref{gapeq}) and (\ref{anomaly}).  
Note that the anomaly (\ref{anomaly}) is crucial to give the term proportional to $\lambda(x)$ in the energy density, after using the gap equation.

It turns out that 
$ {\cal E}_0(x)  $  is  a constant.
The space derivative of $ {\cal E}_0(x)  $ is:
\bea     \frac {d {\cal E}_{0} (x) }{dx}  &=&   N    \sum_{n=1}^\infty 
\left( \omega_n
 f_n  f'_n+\frac{1}{\omega_n}   f'_n  f''_n \right)  e^{-\epsilon  \omega_n}    + 2  \sigma' \sigma''\; \nonumber \\
 &=&    N    \sum_{n=1}^\infty 
\left( \omega_n
 f_n  f'_n+\frac{1}{\omega_n}   f'_n  (\lambda - \omega_n^2) f_n \right)  e^{-\epsilon  \omega_n}    + 2  \sigma' ( \lambda \sigma ) \; \nonumber \\
&=& \lambda(x) \left[     N    \sum_{n=1}^\infty    \frac{1}{\omega_n}  f_n  f'_n   e^{-\epsilon  \omega_n}    +  2  \sigma  \sigma'     \right] \;.
\eea
Noting that the expression in the square bracket above is the derivative of the gap equation  (\ref{gapeq}),  we conclude that
\beq
 \frac {d {\cal E}_{0} (x) }{dx}  &=&  0 \ .
\eeq
We thus find that the energy density is  
\be    {\cal E}(x, \Lambda, L)=   {\cal E}_0(\Lambda, L)  +   \frac{N}{2\pi}  \lambda(x, \Lambda, L)  +  {\cal E}_{\rm uv}\;,  \label{result0}
\ee
where the only dependence on $x$ is through the  function $\lambda(x)$.  
%Eq.~(\ref{result0})   is the first significant result of the present paper.  

Let us study the   $L \to \infty$   limit of this expression.  
The value of the constant part of the energy density,  ${\cal E}_{0}$,  in the  $L\to \infty$  limit  may be calculated
by noting that $\lambda(x) \to \Lambda^2$ and  the fact that the spectrum is exactly known in that limit. 
See Appendix \ref{sec:infiniteL}.  The result is 
\be    {\cal E}_0 (\Lambda,L=\infty)    = \frac{N}{\pi \epsilon^2}   -  \frac{N  \Lambda^2  }{  4  \pi } \;.
\label{divergence}   \ee
This shows that the energy density of the system, after the standard regularization and renormalization of the coupling constant has been  made  to render   the 
gap equation finite,  still contains a quadratic divergence. This is a little similar to the vacuum density in QCD:  the theory can be renormalized
 and all physical quantities can be calculated order by order in perturbation theory,   but the  vacuum energy density  (a contribution to the cosmological constant)
 is still divergent, and requires a further subtraction. 
The result  (\ref{divergence}) however  suggests that we take the vacuum energy subtraction constant simply  as  
\be        {\cal E}_{\rm uv}      = -  \frac{N}{\pi \epsilon^2} \;, \label{subtract}
\ee
 and the constant part of the energy density is thus
\be    {\cal E}_{0} (\Lambda, L=\infty) + {\cal E}_{\rm uv} =    -  \frac{N  \Lambda^2  }{  4  \pi } \;.
\ee
As \footnote{This follows both from the numerical results given in
  \cite{BKO}, and analytical calculations such as in the Appendices, as well as from the general observation that the generalized gap equation itself  reduces to the known  equation of the standard 2D ${\mathbb CP}^{N-1}$ model.  
}
 \be \lambda(x, \Lambda, L)    \to  \Lambda^2,    \qquad    \forall x, \quad  x \ne \pm \frac{L}{2}, \qquad  L \to \infty\;,   \label{lambdaL}    \ee
one finds that the total  energy density approaches a constant
\be     {\cal E}(x, \Lambda, L=\infty)   =       \frac{N}{ 4  \pi }  \Lambda^2 \; ,  \label{analogous}
\ee
at any fixed finite  $x$.    This gives the quantum corrections \footnote{Our system can be  interpreted either as 
a low-energy  effective action of the monopole-vortex soliton complex, 
or  as just an ad hoc  ${\mathbb CP}^{N-1}$ model  defined on a finite worldstrip. In the first case,  the vortex energy scale (or  the vortex classical tension), plays the role of the UV cutoff. 
One is interested in the effects of the  quantum fluctuations of the orientational zeromodes at lower energies, i.e., at length scales larger than the vortex width.  In the latter case, a UV cutoff is introduced 
to renormalize the gap equation (the coupling constant renormalization) and to renormalize the vacuum energy.   From this latter point of view 
(\ref{analogous}) is analogous to the vacuum energy density in QCD. }
due to the fluctuations of the $n_i$ fields to the classical  "tension" of the vortex,  $\xi\,$,   where   $\Lambda^2  \ll  \xi\,.$ 
This result is in agreement with the one \cite{Monin:2015xwa}, found in a finite worldstrip ${\mathbb CP}^{N-1}$ model with periodic boundary conditions. 

As we shall see in the next section, and as can be verified by a WKB analysis done  in Appendix \ref{sec:WKB},  the divergence in the energy density  remains purely quadratic: the only subtraction needed is 
(\ref{subtract}),  in the case of finite  ($L$)  string also.   No linear or logarithmic divergences are present.  This is reasonable as the divergences due to the fluctuations of the $n_i$ fields
is a short-distance effect, local in $x$, and  cannot depend on the presence of  the boundaries.

\section{Large but finite  $L$: an Ansatz and analytic calculation   \label{sec:Analytic} }

We now study the corrections to (\ref{lambdaL}) and (\ref{analogous}) (and $\sigma(0) =0$)  for  large but finite  $L$.   
In order to do that, it is clearly necessary to analyze the large-$L$   behavior of (the solution of)  the gap equation, (\ref{gapeq}), (\ref{lambda}) itself.   
To start with,   $\lambda(x, \Lambda, L)$ can be  approximately  taken to be a constant for $|x|\ll L/2$;  we parametrize its asymptotic approach to $\Lambda^2$ as  
\begin{eqnarray}
 \lambda(0, \Lambda, L) =  \tilde \Lambda^2
\equiv  \Lambda^2 e^{2a},\qquad \lim_{L\to \infty}a(\Lambda L) =0\;. \label{tostart}   \end{eqnarray}
The factor $a$ represents  the nonlocal effect due to the boundary condition. 
To estimate this, consider the propagator,  Eq.~(\ref{thetwo}).   
For $x,x' \sim 0 $ (i.e., far from the boundaries), it satisfies locally
\begin{eqnarray}
 \left(\partial_\epsilon^2+\partial_x^2 -\tilde \Lambda^2\right)
D(x,\epsilon;x',0)\sim \delta(\epsilon)\delta(x-x')\;,   \label{byusing}
\end{eqnarray}
thus  its solution can be assumed to have the form
\begin{eqnarray}
 D(x,\epsilon;x',0)  &\sim&    \frac1{2\pi} 
K_0\big(\tilde \Lambda \sqrt{(x-x')^2+\epsilon^2}\big)       \nn  
&&    
-\frac{A}{2\pi} \left(K_0\big(\tilde \Lambda \sqrt{(x+x' +L)^2+\epsilon^2}\big)
+K_0\big(\tilde \Lambda \sqrt{ (L-x-x')^2+\epsilon^2}\big)\right)   \nn  
&& +\cdots   \label{usethis}
\end{eqnarray}
with an unknown constant $A=A(\Lambda L)$.\footnote{$K_0$ and $K_{1,2}$ below are the modified Bessel functions of the second kind.}  The subleading  terms in the second line come from 
the nearest mirror poles in  Eq.~(\ref{Dequation}).\footnote{
Strictly speaking,  far from the boundaries,  the positions of the mirror poles might also be  effectively  shifted.
Dominant effects of their shift, however,  just rescale factor $A$ and we omit such shifts here for simplicity.    
}
  It is expected that  $A=A(\Lambda L)\sim \mathcal{O}(1)$, but due to the effects of the boundaries where $\lambda(x)$ is non constant and singular \cite{BKO}
it will not coincide  with the exact value   
$A=1$  (for the Dirichlet boundary condition at $x=\pm L/2$)  or  $A=-1$   (for the Neumann boundary condition)   (cfr.  Eq.~(\ref{both2}))  \cite{BKO}.

The gap equation (\ref{gapeq})  then  yields near $x=0$
\begin{eqnarray}
 \sigma^2(x)&\sim&  \frac{N}{2\pi}\log \frac{\tilde \Lambda }{\Lambda } 
+\frac{NA}{2\pi} \left(K_0\big(2\tilde \Lambda (x+L/2) \big)+K_0\big(2\tilde \Lambda (L/2 -x)\big)\right)\nn
&\simeq  & \frac{N}{2\pi}a+\frac{NA}{\pi} K_0(\tilde \Lambda L)
+\frac{2NA \tilde \Lambda^2}{\pi}K_0''(\tilde \Lambda L) \, 
x^2+\cdots
\end{eqnarray}
Equation (\ref{lambda})  gives   $\lambda(x)$  near $x=0$    ($\sigma'\simeq 0$):
\begin{eqnarray}
 \lambda(x)= \frac{\sigma''(x)}{\sigma(x)}\sim         \frac{(\sigma(x)^2)''}{  2 \,   \sigma(x)^2}  \sim
\tilde \Lambda^2  
\frac{2 AK_0''}{a/2+A K_0}\;.
\end{eqnarray}
By requiring the consistency  of this  with the initial Ansatz (\ref{tostart}) and making use of the identity
\be  K_0(x)+K_2(x)=2K_0''(x)\;, \ee 
 one finds that  
\begin{eqnarray}
 a= 2 A\, K_2(\tilde \Lambda L) =  2 A\, K_2(\Lambda L  e^a  )  \simeq      2 A\, K_2( \Lambda L)    \; ,
\end{eqnarray}
and 
\be      \tilde \Lambda^2
\equiv  \Lambda^2 e^{2a}  \simeq    \Lambda^2  ( 1+ 2a) \sim    \Lambda^2   \left(1+4 A K_2(\Lambda L)\right)\;.  \label{Lamdtild}
\ee
Thus  one finds at  large $L \gg \tfrac{1}{\Lambda}$   that, around $x=0$,
\begin{eqnarray} \lambda(x, \Lambda, L) &\sim&  \lambda(0, \Lambda, L) \sim
\Lambda^2 \left(1+4 A K_2(\Lambda L)\right),\nn
  \sigma^2(x, \Lambda, L) &\sim& 
\frac{2NA}{\pi}K_0''(\Lambda L)\left(1+\Lambda^2  x^2   \right) \;.  \label{reflect1}
\end{eqnarray}

With these results in hand, one can now calculate the asymptotic behavior of the energy density   itself:
\be    {\cal E}(x, \Lambda, L)=   {\cal E}_0(\Lambda, L)  +   \frac{N}{2\pi}  \lambda(x, \Lambda, L)  +  {\cal E}_{\rm uv}\;.  \label{result1}
\ee
It turns out under the same approximation (\ref{usethis}) that the constant part of the energy density, ${\cal E}_0(\Lambda, L)$, is given at finite large $L$  by
\be 
 {\cal E}_0(\Lambda, L) 
\sim     \frac{N}{\pi\epsilon^2}
-\frac{N \Lambda^2}{4\pi}
-\frac{N A \Lambda^2}{\pi}\left(K_0(\Lambda L) +K_2( \Lambda L)\right)  +{\cal O}(e^{-2\Lambda L})\;.
\ee
The derivation of this result is given in Appendix~\ref{ConstEnergy}.    Note that the  divergence  in the energy density is just the  purely quadratic 
one, $ \frac{N}{\pi\epsilon^2}$, the same as in the $L \to \infty$ case  discussed in the previous section.  This is correct, as  the divergences arise from the UV fluctuations of the
$n_i$ fields which is a local effect, independent of the boundaries, or of the value of $L$. 

Finally one finds,  by adding the $\lambda(0, \Lambda, L)$ term and by
making the same subtraction as before, viz.~(\ref{subtract}),    
\begin{eqnarray}
 {\cal E}(0, \Lambda, L)&=&\frac{N}{2\pi}\lambda(0, \Lambda, L)+{\cal
  E}_0(\Lambda, L)     +  {\cal E}_{\rm uv}   \nn
&\sim& \frac{N\Lambda^2}{4\pi}
+\frac{NA \Lambda^2 }{\pi}(K_2(\Lambda L)-K_0(\Lambda L))+{\cal O}(e^{-2\Lambda L})\nn
 &=&  \frac{N \Lambda^2 }{4\pi}  + 
\frac{2N A \Lambda }{\pi L} K_1(\Lambda L)+{\cal O}(e^{-2\Lambda L})  \;,    \label{reflect2}
\end{eqnarray}
where another  identity  
\be  K_2(x)-K_0(x)=2K_1(x)/x \; , \ee
 has been used.

To conclude, we find that the approach to the asymptotic value of the energy density   is  exponential: no pure power 
corrections in  $1/L$  (i.e.~the L\"uscher  term)  are present.  This is perfectly consistent with the general result found in \cite{BKO} that our system has a unique phase, 
which smoothly matches -- in the large $L$ limit --  the ``confinement
phase'' of the standard 2D ${\mathbb CP}^{N-1}$ model.  All $n_i$
($i\ne 1$)   fields gain a dynamically generated mass $\sim \Lambda$;  at the same time $\sigma \sim 0$
except at the boundaries. In other words, no dynamical breaking of the isometry group $SU(N)$  takes pace. No Nambu-Goldstone modes associated with the internal, orientational modes  are generated.   The absence of a long-range correlation in  the large-$L$ corrections in Eqs.~(\ref{reflect1}), (\ref{reflect2}) is a simple reflection of this fact.

\section{Numerical results}  
\label{six}

Due to the quadratic divergence present in the sum
(\ref{EdensityConst})  the numerical calculation turns out to present
quite a bit of a challenge.  Any tiny errors in the eigenmodes and in the energy
levels will introduce   linear or logarithmic divergences in the sum,  and the finite answer  for 
${\cal E}(x, \Lambda, L)$ one gets (including its dependence on $x$, $L$ and $\Lambda$) 
depends on how these fake divergences are appropriately subtracted, together with the genuine quadratic divergences.   Because of this,  even the best results so far do not have 
a precision comparable to  the solution of the generalized gap
equation discussed in Section~\ref{sec:gapeq}.

\begin{figure}[!htp]
\begin{center}
\includegraphics[width=3.5in]{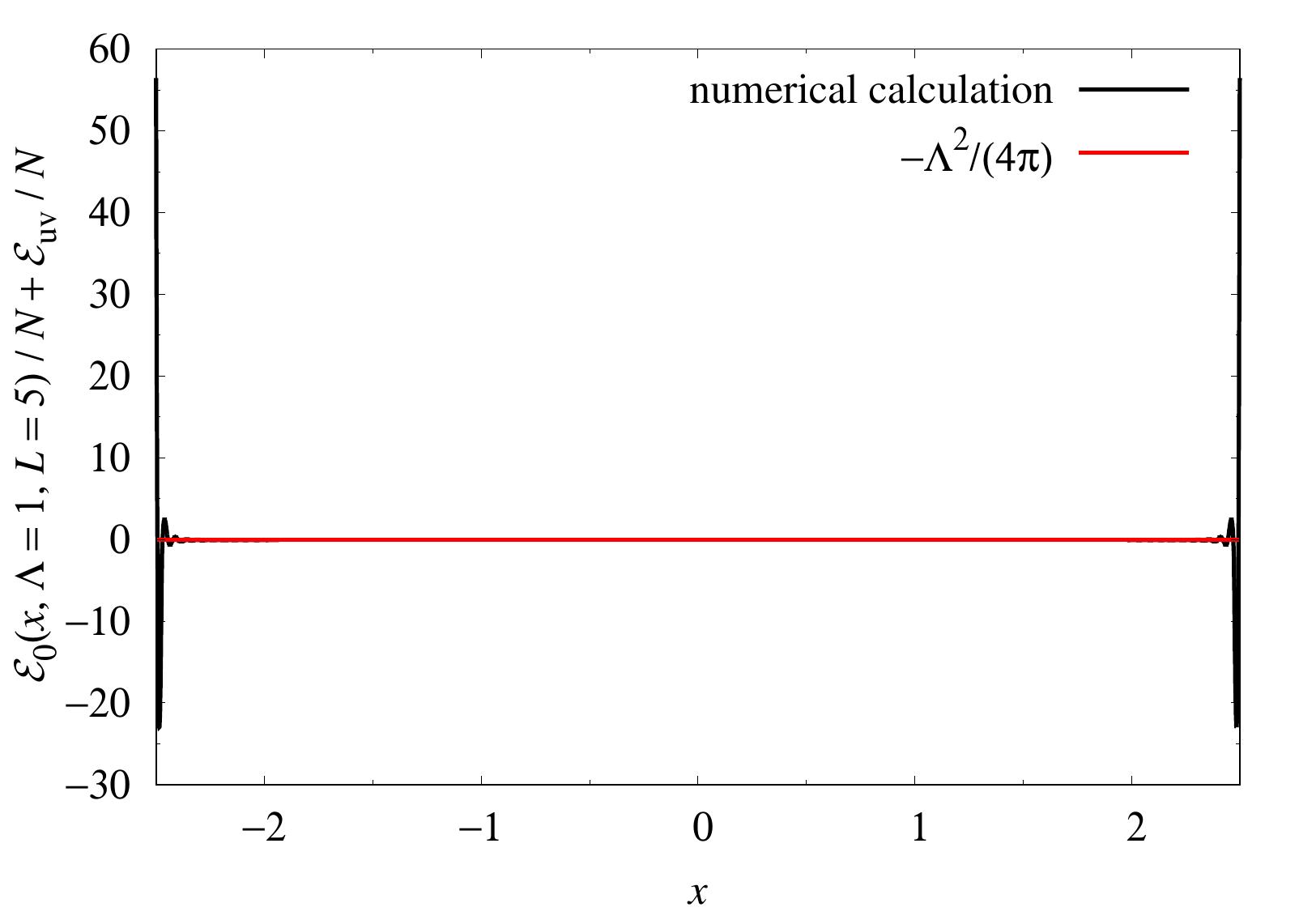}\vspace{1em}\\
\caption{Verification of the constancy of
  $({\cal E}_0+{\cal E}_{\rm uv})$ with respect to
  $x$. $\Lambda=1$, $L=5$ in this figure. }
\label{E0const}
\end{center}
\end{figure}

\begin{figure}[!htp]
\begin{minipage}{0.49\linewidth}
\begin{center}
\includegraphics[width=3in]{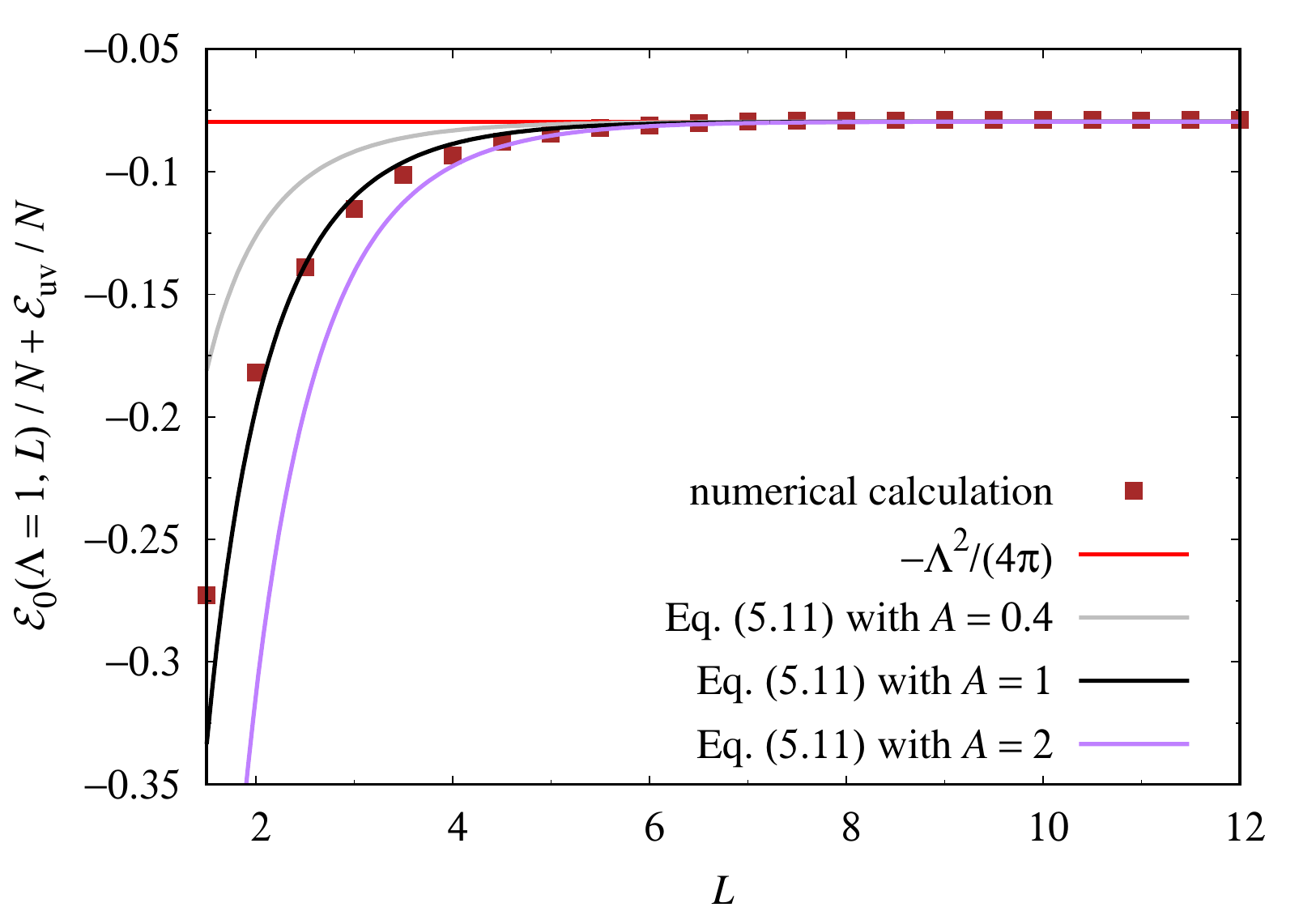}\vspace{1em}\\
\caption{The constant part of the energy density,
  ${\cal E}_0+{\cal E}_{\rm uv}$, is plotted as a function of $L$ for
  $\Lambda=1$ fixed.  }
\label{E0}
\end{center}
\end{minipage}\ \ \
\begin{minipage}{0.49\linewidth}
\begin{center}
\includegraphics[width=3in]{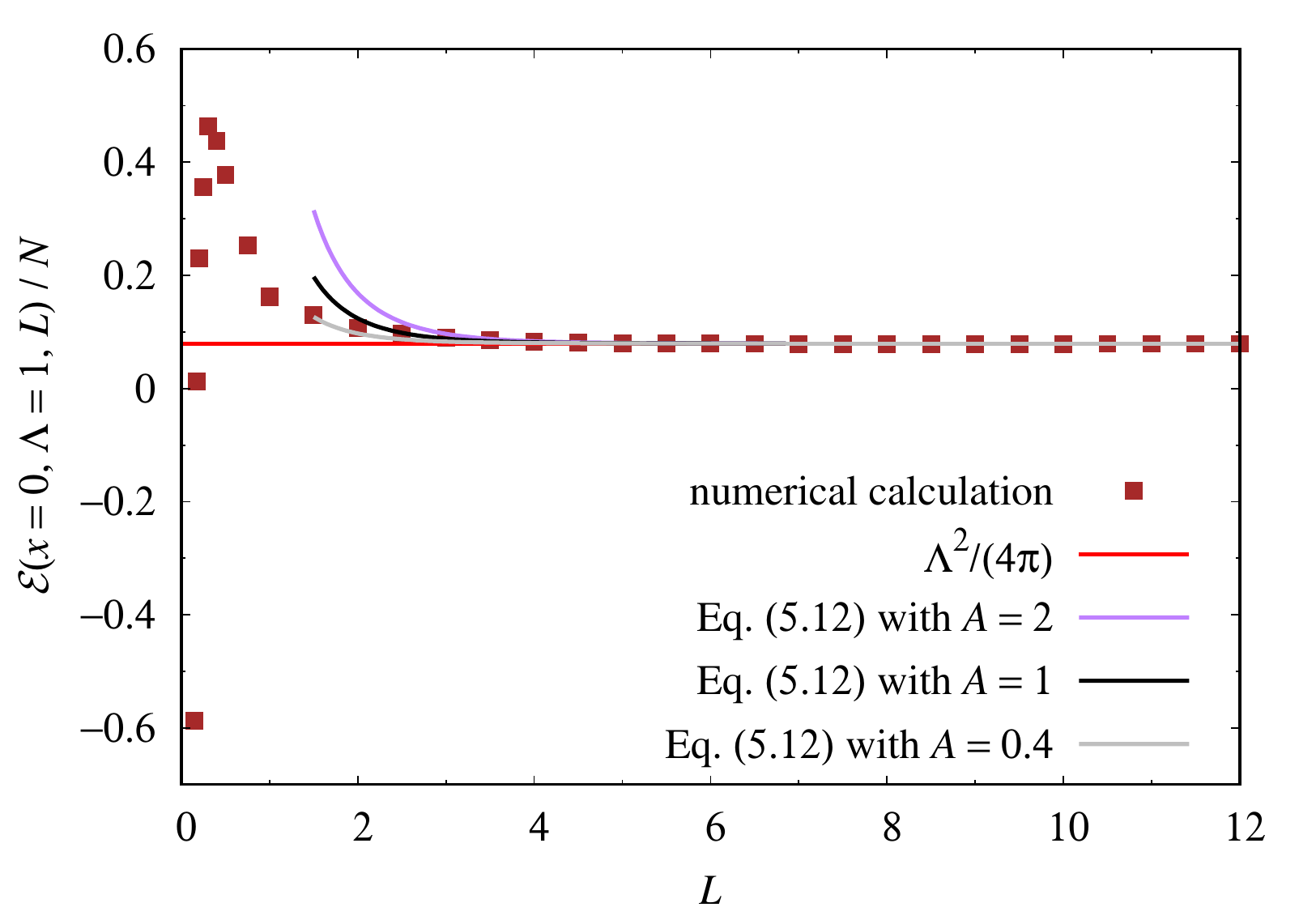}\vspace{1em}\\
\caption{The total energy density at the midpoint, ${\cal E}$, is plotted against $L$ for $\Lambda=1$ fixed. }
\label{E}
\end{center}
\end{minipage}
\end{figure}

\begin{figure}[!htp]
\begin{minipage}{0.49\linewidth}
\begin{center}
\includegraphics[width=3in]{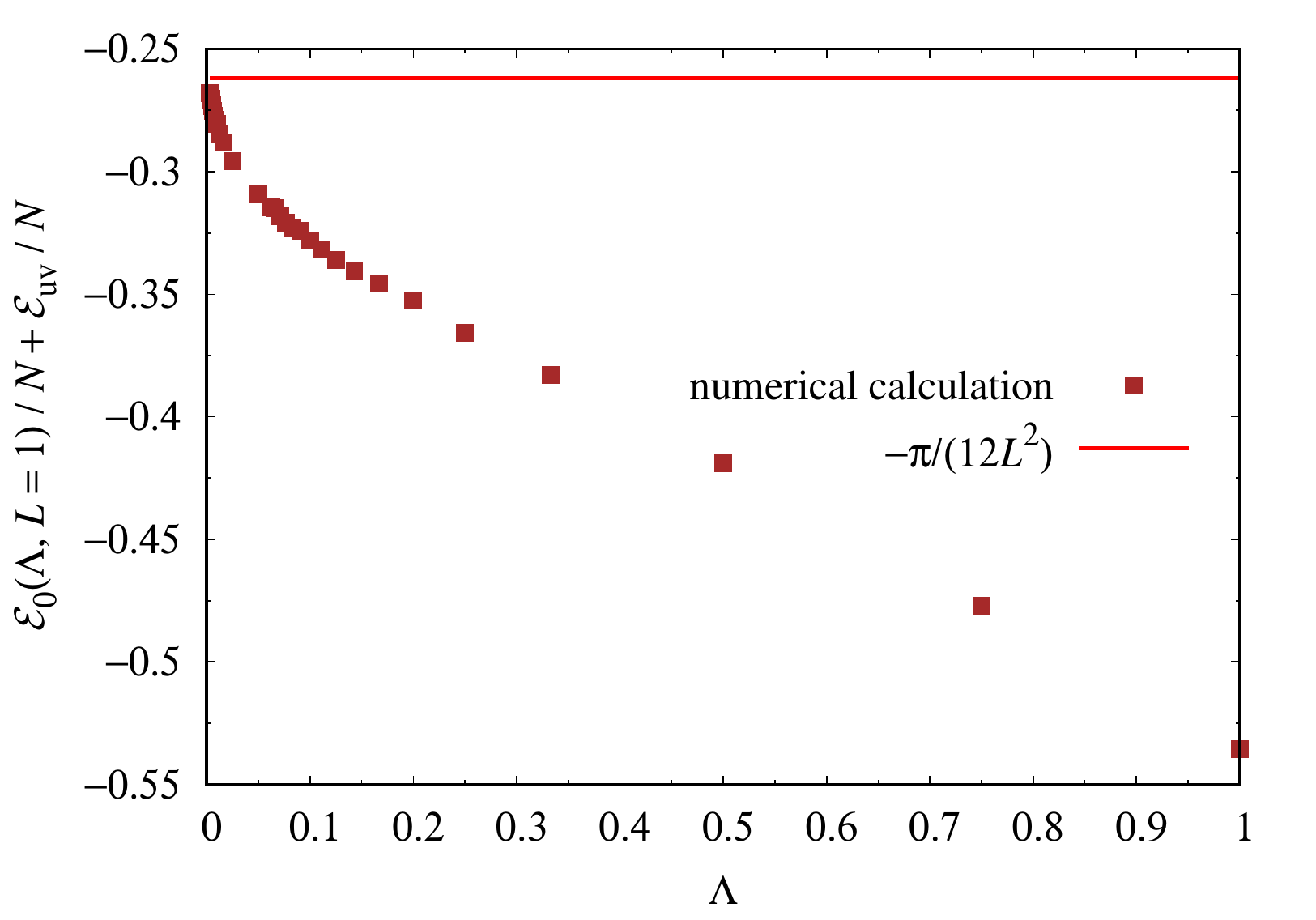}\vspace{1em}\\
\caption{The constant part of the energy density,
  ${\cal E}_0+{\cal E}_{\rm uv}$, is plotted as a function of $\Lambda$ for $L=1$ fixed.  }
\label{E0bis}
\end{center}
\end{minipage}\ \ \
\begin{minipage}{0.49\linewidth}
\begin{center}
\includegraphics[width=3in]{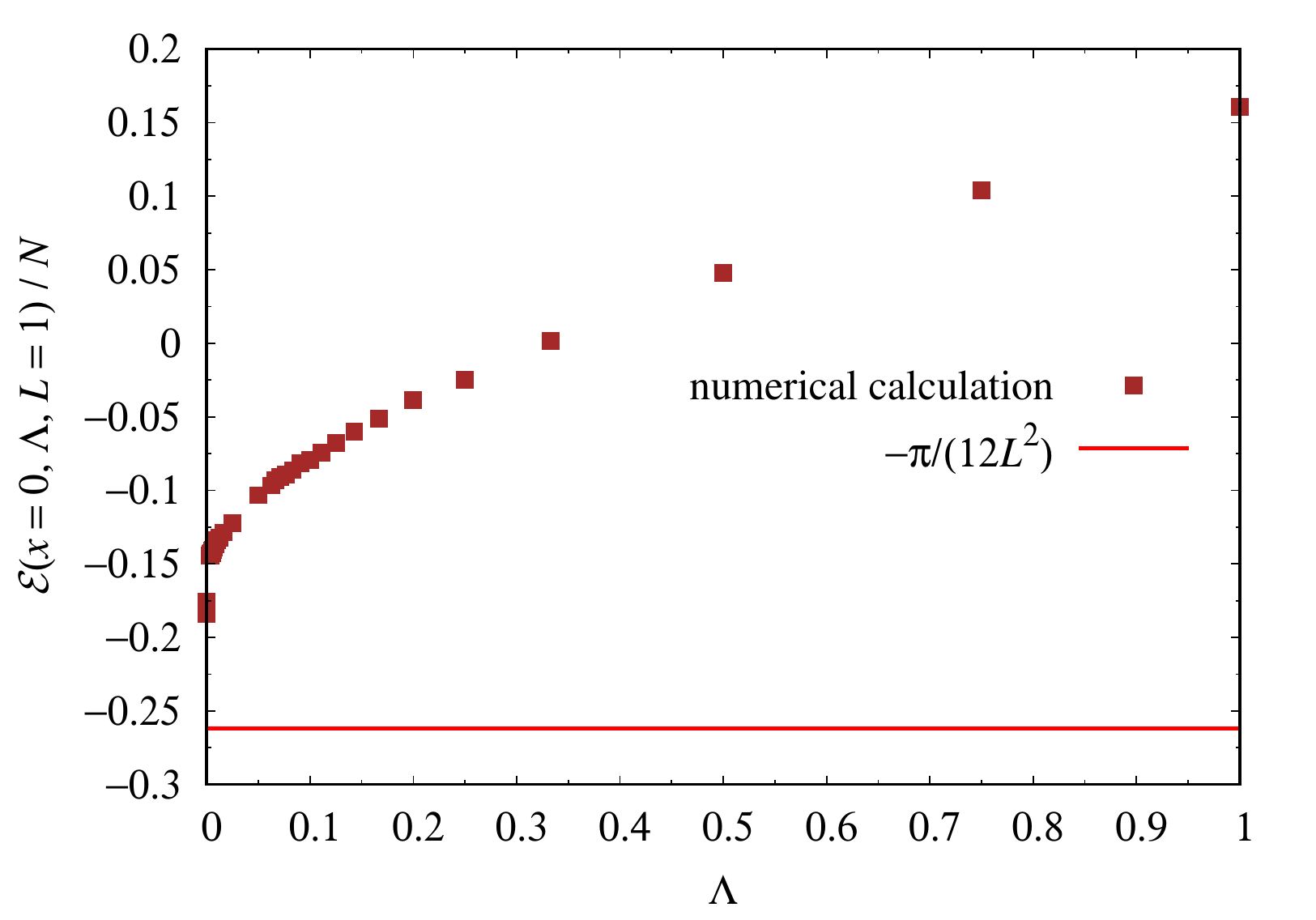}\vspace{1em}\\
\caption{The total energy density at the midpoint, ${\cal E}$, is plotted against $\Lambda$ for $L=1$ fixed. }
\label{Ebis}
\end{center}
\end{minipage}
\end{figure}

The check of the  constancy of  ${\cal E}_0$, is shown in
Fig.~\ref{E0const}. Note that it  is  found indeed to be constant 
everywhere, 
including values of $x$  very close to the boundaries,  where the distances from the latter  are much smaller than $1/\Lambda$. 
Fig.~\ref{E0}  shows  the value of  ${\cal E}_0$ as a function of $L$ ($\Lambda=1$).
The total energy density, including the $N\lambda(x)/2\pi$ term,  calculated at the midpoint, $x=0$, is shown in Fig.~\ref{E}.
The numerical results are nicely consistent with the exact result at
$L=\infty$, and with the analytic behavior  for large but finite $L$, found  in the previous section, with $A \sim 1$.  
 ${\cal E}_0$ and  ${\cal E}(0, \Lambda, L)$  are plotted against $\Lambda$ at fixed $L$,  in  Fig.~\ref{E0bis} and in Fig.~\ref{Ebis}.
In particular we see that as $\Lambda \to 0$ the energy density
converges, although quite slowly (i.e.~logarithmically),  to the free-field value $-\pi/12 L^2$.

\newpage
\section{Boundary divergence and the Casimir force}  
\label{seven}

The energy density 
\be    {\cal E}(x, \Lambda, L)=   {\cal E}_0(\Lambda, L)  +
\frac{N}{2\pi}  \lambda(x, \Lambda, L) + {\cal E}_{\rm uv} \;,  \label{result2}
\ee
after renormalization is  a finite function of $x, \Lambda, L$.
When integrated over $x\in \left[-\frac{L}{2}, \frac{L}{2}\right]$,  it  gives the total energy of the string
\bea   E(\Lambda, L) &=&   \int_{-L/2}^{L/2}  dx \;   {\cal E}(x, \Lambda, L) = 
\int_{-L/2}^{L/2}  dx \, \left[  {\cal E}_0(\Lambda, L) +  \frac{\lambda(x, \Lambda, L)}{2\pi}\right] \nonumber \\
&=&  L\,  {\cal E}_0(\Lambda, L)    +     \int_0^{L/2} dx \;  \frac{\tlambda(x, \Lambda, L)}{\pi}  \;, \label{totenergy}
\eea
where we introduced the mass gap function defined on the interval
$[0,L]$,
\be
\tlambda(x,\Lambda, L) \equiv \lambda(x-L/2,\Lambda,L)\,.
\ee
The Casimir force is defined as \footnote{With this definition a positive ({\it vis a vis},  a negative)  $F$ corresponds to an attractive  (repulsive) force.}
\be   F= \frac{\de E(\Lambda, L)}{\de L} \;.    \label{Casimir} 
\ee

Before analyzing the behavior of $F$ for various values of $L$, 
let us note that the second term in Eq.~(\ref{totenergy})
 gives rise to a new divergence in the integrated energy  due to the singular behavior near the boundaries, e.g., near $x=0$,  
\be
\tlambda(x) \sim \frac{1}{x^2\log 1/(x\Lambda)}\;, \qquad
L \gg  1/\Lambda  \gg  x \;.
\label{leaddiv}
\ee
 As $\tlambda(x)$  quickly approaches the constant value  $\Lambda^2$  beyond $x \simeq  1/\Lambda$, for sufficiently large $L$,  the divergent part can be extracted by considering the finite  integral
\be    E_1  = \frac{N}{2\pi}\int_{\epsilon} dx\;    (\tlambda(x)- \Lambda^2) \sim   \frac{N}{2\pi}    \frac{1}{\epsilon \log{1/\epsilon  \Lambda} }\simeq    \frac{N}{2\pi \epsilon  }   \frac{N g(\epsilon)^2}{8 \pi^2}  \;,  \label{FITWITH}
\ee
where a UV cutoff  ($x=\epsilon$) has been introduced.  Similarly $E_2$ for the contribution from  the right boundary.
$E_{1}$  ($E_{2}$) is an energy concentrated at the left (right) boundary:  it can be  interpreted as the quantum corrections to the monopole  (antimonopole) mass, 
due to the $n_i$ field fluctuations (the factor $N$).   $E_{1,2}$  can be subtracted (i.e., compensated with the bare mass terms)   from the total  energy,   leaving finite, renormalized monopole masses. They do not affect the discussion on the $L$ dependent Casimir effect below. 
%}

The Casimir force can be rewritten, by differentiating Eq.~(\ref{totenergy}),   as
\bea   F &=&  {\cal E}_0(\Lambda, L) + L\, \frac{\de{\cal E}_0(\Lambda, L)  }{\de L}   +     \frac {\lambda(0, \Lambda, L)}{2\pi} + \frac{1}{\pi} \int_0^{L/2} dx \,   \frac{\de\tlambda(x, \Lambda, L)}{\de L}  \nonumber \\
&=&  {\cal E}(0, \Lambda, L) + L\, \frac{\de{\cal E}_0(\Lambda, L)  }{\de L}  + \frac{1}{\pi} \int_0^{L/2} dx \,   \frac{\de\tlambda(x, \Lambda, L)}{\de L} \;.\label{casforce}
\eea
Note that this is a finite quantity since it is not sensitive to the leading divergence of $\tlambda$, viz.~(\ref{leaddiv}).

At very small $L$ (i.e.~$L\Lambda\ll1$) one expects the dominant effect to come  from the second term of (\ref{casforce})   (see Fig.~\ref{Ebis}):
\be     F \simeq  L\, \frac{\de{\cal E}_0(\Lambda, L)  }{\de L}
\simeq \frac{N\pi}{6 L^2} \;:
\ee
 it is an attractive free-field Casimir force.

At intermediate values,  $L\sim \mathcal{O}(1/\Lambda)$, instead, we find that the force is dominated by the third term in  (\ref{casforce}).  The strong decrease ($\sim 1/ L^2 \log (1/L)$)  of the mass gap $\tilde {\lambda}(x)$ with $L$ for all $x$ (see e.g., Eq.~(\ref{decrease})) cannot be compensated by a linear effect of integration, therefore  there is an effective repulsive force at work.  The total energy of the system is lowered when the space interval $L$ gets larger.  

At sufficiently large $L$,  $L  \gg1/\Lambda$, where the $2D$ regime sets in,   the leading contribution comes from the  first term in (\ref{casforce}):
\be   F\simeq  {\cal E}(0, \Lambda, L)  \simeq  \frac{N \Lambda^2
}{4\pi} \;,
\ee
corresponding to an approximately constant string tension (Eq.~(\ref{analogous})).  An external observer who attempts to pull the boundaries further apart will experience an attractive, constant force countering her/him.

A precise numerical verification of this nontrivial behavior of the force turned out to be 
exceedingly difficult because of the singular behavior of $\lambda(x)$
near the boundaries.  Our preliminary result (not shown)  however clearly confirms the change from a repulsive regime at $L =  \mathcal{O}(1/\Lambda)$ to an attractive force at  
 $L \gg 1 / \Lambda$.

\section{Conclusion} 
\label{eight}

In this paper we have examined the  energy density  function  ${\cal E}(x, \Lambda, L)$  of the  large-$N$  $\mathbb CP^{N-1}$ sigma model on a finite string, defined with the Dirichlet boundary conditions. We find that it is a sum of  two terms, the first  expressed as a sum over fluctuation modes, which turns out to be constant in $x$, and the second term proportional to the mass gap
$\lambda(x)$. The only $x$ dependence arises from the second.  The first term is quadratically divergent, analogous to the  vacuum energy in QCD. 
The $L$-dependence of ${\cal E}(x, \Lambda, L)$  at fixed $\Lambda$
shows that the effect of the boundaries are limited to their vicinity of width $\sim 1/\Lambda$:  the system approaches quickly the standard 2D large-$N$  $\mathbb CP^{N-1}$ sigma model,
with dynamical generation  of the mass gap, and with no dynamical breaking of the  isometry group $SU(N)$.  In the small $L\Lambda$ limit, the system approaches
the classical  weakly-coupled $\mathbb CP^{N-1}$  model, as appropriate for the Dirichlet boundary conditions.   

 The approach to  the limit 
$ {\cal E}(x, \Lambda, L=\infty)   =       \tfrac{N}{ 4  \pi }  \Lambda^2 $   is  found to be  purely  exponential: no   power 
corrections in  $1/L$  such as   the L\"uscher  term  are present.  This is perfectly consistent with the general result found in \cite{BKO} that our system has a unique phase, 
which smoothly matches the ``confinement phase'' in the large-$L$ limit
of the standard 2D  ${\mathbb CP}^{N-1}$ model.  All $n_i$ ($i\ne 1$)
fields gain a dynamically generated mass $\sim \Lambda$;  at the same time $\sigma \sim 0$
except at the boundaries. In other words, no dynamical breaking of the isometry group $SU(N)$  takes pace, and no associated Nambu-Goldstone modes are generated.   The absence of the  long-range correlation reflects this fact.

Recently a  paper appeared \cite{Nitta:2017uog} in which
 some  analytical large-$N$  $\mathbb CP^{N-1}$ sigma model solutions
 for  inhomogeneous condensates are presented by a mapping to the Gross-Neveu model \cite{Flachi:2017cdo}. These solutions correspond to 
 {\it periodic} boundary conditions and, as far as we can see, 
  none of the solutions proposed there correspond to our system defined with the Dirichlet boundary conditions. It would certainly be very interesting if our type of solution could be found analytically with developments of these techniques in the future.

\subsection*{Acknowledgments}

We thank Jarah Evslin, Muneto Nitta, Misha Shifman and Ryosuke Yoshii
for discussions. 
The work of S.~B.~is funded by the grant ``Rientro dei Cervelli Rita
Levi Montalcini'' of the Italian government.
The work of S.~B.~G.~was supported by the National Natural Science
Foundation of China (Grant No.~11675223).
K.~O.~is supported by the Ministry of Education, Culture, Sports,
Science (MEXT)-Supported Program for the Strategic Research Foundation 
at Private Universities ``Topological Science'' (Grant No.~S1511006)
and by the Japan Society for the Promotion of Science (JSPS)
Grant-in-Aid for Scientific Research (KAKENHI Grant No.~16H03984). 
The present research work is supported by the INFN special research
project grant, GAST (``Gauge and String Theories'').

When this paper was being prepared for submission we were informed by Muneto Nitta and Ryosuke Yoshii of their paper 
\cite{NittaYoshiiNew}, which deals with similar problems as ours, and with some overlap. 
 The boundary behavior for the gap function
and some other qualitative aspects of their solutions are different from ours.  Also, another paper  \cite{Pavshinkin}  just appeared,
 discussing a Grassmannian sigma model on finite-width world sheets.

\subsection*{Comments on Ref. \cite{NittaYoshiiNew}}

In Ref.~\cite{NittaYoshiiNew}, submitted to the ArXiv on the same day
as ours, the same system is analyzed in a different approach, and the
authors there claim to find analytic solutions for mass gap function
$\lambda(x)$ and for $\sigma(x)$, both in confinement and in Higgs
phases. 

By imposing Dirichlet boundary conditions and solving the generalized
gap equations, we instead find the solutions (e.g., illustrated in
Fig.~\ref{fig1}) in a unique phase with mass gap for all values of
$L$, which smoothly approaches the well-known solution in the infinite
$L$  limit (the standard $2D$ ${\mathbbm C}P^{N-1}$ model). Our
solutions are moreover consistent with the classical
${\mathbbm C}P^{N-1}$ model in the $L \ll 1/\Lambda$ limit as
discussed in Sec.~\ref{section:near}.

It is possible that, if a Higgs-like solution (as in
Fig.~1(b) of \cite{NittaYoshiiNew}) would exist, it represents
an unstable solution, whereas our procedure necessarily picks up the 
stable solution, and that actually the confinement-type solution is
always the stable one. 

However, we find it difficult to make a proper comparison, as the
renormalization of the gap equation and the generation of the mass
scale $\Lambda$ are not explained in \cite{NittaYoshiiNew}. 

The boundary behavior of the mass gap function and the field $\sigma$
given in \cite{NittaYoshiiNew} is powerlike, whereas the logarithmic
behavior found by us reflects the situation characteristic of a
finite-space-width system.  The system must compromise between the
$2D$ physics at $L \ge  1/\Lambda$ - the divergences of the $n_i$
field fluctuations and the generation of the mass scale $\Lambda$ -
and the classical limit to which the model must reduce correctly in
the $L \ll 1/\Lambda$ region, as explained in Sec.~\ref{section:near}.

As the physical values of $L$ (the space width) are not given in
reference to $1/\Lambda$ in \cite{NittaYoshiiNew}, in contrast to what
is done in the present paper, it is not clear to us which physical
values of $L$ their solutions in Fig.~1(b) or Fig.~1(a) refer to, for
instance. 

As a consequence, it is unclear how and when (at which value of $L$)
the Higgs phase vacuum disappears, as $L$ is increased. Or, vice
versa, at which $L$, if $L$ decreases toward zero, the Higgs vacuum
takes over, if it does at all. The authors of \cite{NittaYoshiiNew} do
not give the criteria to decide which solutions should be chosen at
any given $L$.  As far as we can see, the analysis of the vacuum
energy density, as made in the present paper, has not been done yet
there.

\appendix

\section{ Propagator $D(x,\tau;x',\tau') $\label{app:prop}}
\subsection{ Exact forms with $\lambda(x)=m^2$ }
When $\lambda(x)=m^2$, 
Eq.~(\ref{Dequation}) can be easily solved by
\bea       D(x,\t; x',\t')  &=& \sum_{n\in Z}    \frac{1}{2\pi}  K_0\big(m \sqrt{ (x- x^{\prime} + 2n L)^2 +  (\t- \t^{\prime})^2} \big)  \nonumber \\
&&  -    \sum_{n\in Z}    \frac{1}{2\pi}  K_0\big(m \sqrt{ (x+ x^{\prime} + (2n+1) L)^2 +  (\t- \t^{\prime})^2 }\big) \;  ,   \label{both2}
\eea
for the Dirichlet-Dirichlet boundary condition (\ref{DDbc}).
In particular, note that 
\be       D(x,\epsilon; x,0)  =\sum_{n\in Z}    \frac{1}{2\pi}  K_0\big(m \sqrt{ (2n L)^2 +  \epsilon^2}\big)   -    \sum_{n\in Z}    \frac{1}{2\pi}  K_0\big(m \sqrt{ (2 x+ (2n+1) L)^2 +  \epsilon^2}\big) \;.  \label{both3}
\ee   
In the case of the Neumann-Neumann boundary condition (\ref{NNbc}), 
the sign of the last terms of the r.h.s. is flipped.

\subsection{Alternative derivation of the anomalous functional  variation  \label{derivation}}
In terms of the propagator (\ref{thetwo}),  the extra factor  in Eq.~(\ref{anvar}) can be  expressed as
\begin{eqnarray}
{\rm r.h.s.~of~Eq.(\ref{anvar})}=-  \frac{\epsilon    N}{2}   \sum_{n}  f_n(x)^2 e^{-\epsilon \omega_n}
= \epsilon N \frac{\partial }{\partial \epsilon} D(x,\epsilon;x,0).
\end{eqnarray}
Since the UV divergence comes from a short-distance effect,   to extract divergent terms of $D(x,\epsilon,x,0)$,
 it is sufficient to consider the contribution of the nearest poles, 
$\delta(\tau-\tau')\delta(x-x')$ ($n=0$ term) in Eq.~(\ref{Dequation}). Furthermore, at  short distances 
$ \lambda(x) (|x-x'|^2+\epsilon^2) \ll 1$,  the  potential $\lambda(x)$ can be omitted  and thus 
the propagator behaves as one for a massless field in two dimensional space,
\begin{eqnarray}
D(x,\epsilon;x',0) \sim -\frac1{4\pi} \log \left(|x-x'|^2+\epsilon^2 \right)+{\rm regular~terms}.
\end{eqnarray}
With a general potential $\lambda(x)$, therefore,
the divergent part of  ${D}(x,\epsilon;x,0)$   (for $ - \tfrac{L}{2} < x < \tfrac{L}{2} )$   is universal as
\begin{eqnarray}
 { D}(x,\epsilon;x,0)\sim -\frac1{2\pi}\log(\epsilon )
+{\rm regular~terms}\,.
\end{eqnarray}  
In the simplest case with $\lambda(x)=m^2$,  one can easily check this property using Eq.~(\ref{both2}) 
and $K_0(m \epsilon) \sim \log  \tfrac{1}{\epsilon}$.
We find 
\begin{eqnarray}
\lim_{\epsilon \to 0} \epsilon  \frac{\partial}{\partial \epsilon }{ D}
(x,\epsilon; x,0)=-\frac1{2\pi}, \qquad     x\ne \pm \frac{L}2\;,
\end{eqnarray}
which gives the extra  constant term  $  -\frac{N}{2 \pi}$ in the gap equation.   

Similarly, in  a region where
\begin{eqnarray}
\lambda(x) \left( (x-x')^2+\epsilon^2\right) \ll 1, \quad \lambda(x) \left( (x+x'\pm L)^2+\epsilon^2 \right) \ll 1\;,
\end{eqnarray}
the dominant behavior of the propagator is 
\begin{eqnarray}
D(x,\epsilon;x',0) \sim\frac1{4\pi} \log \left(|x+x'\pm L|^2+\epsilon^2 \right)-\frac1{4\pi} \log \left(|x-x'|^2+\epsilon^2 \right)+\cdots \;, \label{both4}
\end{eqnarray}   
where  $(\ldots)$ stands for regular terms. 
For instance,   the two contributions  in (\ref{both4})  exactly cancel each other at the boundaries $x=\pm L/2$,  
consistently with the boundary condition (\ref{DDbc}) for the eigenmodes.

\section{Calculation of $ {\cal E}_{0} (\Lambda, L=\infty) $   \label{sec:infiniteL}}

At large $L$ and at finite $x$,  where $\lambda(x)\sim \Lambda^2$,   one can make an approximation valid at all levels $n$    (simply assume $\lambda=m^2= \Lambda^2$).
Then
\bea     {\cal E}_{0}   &=& \frac N 2   \sum_{n=1}^\infty 
\left( \omega_n
 f_n(x)^2+\frac{1}{\omega_n}   f'_n(x)^2 \right)  e^{-\epsilon  \omega_n}    +(\sigma'(x))^2\;,\label{ddstatesbis}
 \eea
 with
 \begin{eqnarray}
 f_n(x)=\sqrt{\frac2L} \sin\left(\frac{n \pi (x+L/2) }{L} \right) ,\qquad 
\omega_n=\sqrt{\left(\frac{n\pi}L\right)^2+  m^2} \;, \qquad n\geq 1, \quad  n \in \mathbb Z\;.
\end{eqnarray}
As $ {\cal E}_{0} $  has been shown to be a constant, it can be calculated at  any fixed $x$,  for example at the midpoint $x=0$, where $\sigma'=0$:
\be   f_n(L/2)=  \sqrt{\frac2L} \sin\left(\frac{n \pi  }{2} \right)\;, \qquad   f_n'(L/2)=   \sqrt{\frac2L}   \frac{\pi n}{L} \cos\left(\frac{n \pi  }{2} \right)\;.
\ee
In the  $L\to \infty $  limit,  the sum may be replaced by an integral,   by $ \frac{\pi n}{L} \to z$.   Also, let us make a replacement 
\be     e^{-\epsilon \omega_n} \to   e^{-\epsilon \pi n /L}\;, \qquad  {\rm i.e., } \qquad   \omega_n=\sqrt{\left(\frac{n\pi}L\right)^2+  \Lambda^2}  \to  \frac{n\pi}L\;,
     \label{replacement}
\ee
in the exponential damping factor.
One finds 
\bea     {\cal E}_{0}'   &= &\frac{N}{\pi}  \int_0^{\infty}   dz  \, \left(   \sqrt{z^2 + \Lambda^2}  \sin^2 \tfrac{z L}{2}  +   \frac{z^2}{ \sqrt{z^2 + \Lambda^2} } \cos^2 \tfrac{z L}{2} \right)   e^{-\epsilon z}   \nonumber \\
& \sim &     \frac{N  \Lambda^2  }{  2  \pi }    \int_0^{\infty}     dz  \,    \frac{ 2 z^2 +1}{ \sqrt{z^2 + 1} }    e^{-\epsilon z  \Lambda }     \;.     \label{replace}
\eea
Now
\be      \int_0^{\infty}     dz  \,    \frac{ 2 z^2 +1}{ \sqrt{z^2 + 1} }    e^{-\epsilon z  \Lambda }=     \frac{2}{ \epsilon^2 \Lambda^2 } +  \frac{1}{2}  +\mathcal{O}(\epsilon)\;,
\ee
therefore
\be      {\cal E}_{0}'   =    \frac{N}{\pi \epsilon^2}  +    \frac{N  \Lambda^2  }{  4  \pi }  \;.  \label{infiniteL}
\ee

 In going from (\ref{ddstatesbis})  to  (\ref{replace}), however,  we made a replacement   (\ref{replacement})
in the exponential damping factor.  The correction due to this approximation must be taken into account.
The effect of  this replacement can be studied by writing 
\bea    {\cal E}_{0}  &=&   \frac{N    }{  2  \pi }   \int_0^{\infty}     dz \, (\ldots)\,   e^{-\epsilon z   }    e^{-  \epsilon( \omega_n -  z )} \nonumber \\
&=&  \frac{N  }{  2  \pi }   \int_0^{\infty}     dz \, (\ldots)\,   e^{-\epsilon z  }    \, \left[ 1 -   \epsilon( \omega_n -  z ) +\cdots \right]\;. 
\eea
Clearly  the terms of order  $\epsilon^2$ or higher inside $[\ldots ]$   are unimportant, as  the integral in $z$ is finite 
 without the regularizing exponential factor,  or at most logarithmically divergent for the $\epsilon^2$ term.  
The  $\mathcal{O}(\epsilon)$  term in the square bracket  $[\ldots ]$  gives
\be   \epsilon  \,  \frac{N   }{  2  \pi }   \int_0^{\infty}     dz \, (\ldots)\,  ( \omega_n -  z ) \,   e^{-\epsilon z  } \;,
\ee
but
\be     \omega_n -  z    \sim \frac{1}{ 2  z}   \;,
\ee
and 
\be    (\ldots)  \sim   z \;,
\ee  
(see (\ref{replace}))   so  
\be      (\ldots)\,  ( \omega_n -  z ) \sim  1 \;,
\ee
at large $z$:  the integral diverges linearly  as 
\be     \int_0^{\infty}     dz \,  e^{-\epsilon z   }   \sim    \frac{1}{ \epsilon }\,
\ee
so it gives a finite contribution.   It is
\be     -  \epsilon \frac{N \Lambda^3}{2 \pi}  \int_0^{\infty}  dz \frac{ 2 z^2 + 1}{\sqrt{z^2 +1}} \left( \sqrt{z^2 +1} - z  \right)\, e^{-  \epsilon \Lambda z}\;.
\ee
This can be easily calculated to give
\be       -   \frac{N \Lambda^2}{2 \pi}  + \mathcal{O}(\epsilon)  \;.
\ee
This must be added to (\ref{infiniteL}) obtained under  the approximation  (\ref{replacement}): the final answer is 
\be      {\cal E}_{0} (\Lambda, L=\infty)  =    \frac{N}{\pi \epsilon^2}    -  \frac{N  \Lambda^2  }{  4  \pi }  \;.  \label{infiniteLfinal}
\ee

\section{Calculation of  ${\cal E}_0(\Lambda, L) $  \label{ConstEnergy} for $\Lambda L \gg 1$} 

To compute  ${\cal E}_0(\Lambda, L)$   at large  but finite $L$ we observe that the constant part  (evaluated at $x=0$)  of the energy density (\ref{EdensityConst}) can be written as   (see Eq.~(\ref{thetwo}))
\be 
 {\cal E}_0(\Lambda, L) 
=N \left(\frac{\partial^2 }{\partial \epsilon^2 }+\frac{\partial^2
      }{\partial x \partial x'}\right)D(x,\epsilon;x',0)
\Big|_{x,x'=0}+\sigma'(0)^2  \;. \ee
By using (\ref{byusing})  this can be rewritten as 
\be {\cal E}_0(\Lambda, L) 
=  N \left\{\lambda(0, \Lambda, L)+ \frac{\partial }{\partial x }\left(
\frac{\partial }{\partial x' }-\frac{\partial }{\partial x }\right)
\right\}  D(x,\epsilon;x',0)
\Big|_{x,x'=0} \;,
\ee
where we set $\sigma'(0)=0$ by symmetry.  We now use   (\ref{usethis}): 
\begin{eqnarray}
 D(x,\epsilon;x',0)  &\sim&    \frac1{2\pi} 
K_0\big(\tilde \Lambda \sqrt{(x-x')^2+\epsilon^2}\big)   \nn  
&&    
-\frac{A}{2\pi} \left(K_0\big(\tilde \Lambda \sqrt{(x+x' +L)^2+\epsilon^2}\big)
+K_0\big(\tilde \Lambda \sqrt{ (L-x-x')^2+\epsilon^2}\big)\right)   \nn  
&& +\cdots   \label{usethisBis}
\end{eqnarray}
and  (\ref{tostart}), to get 
\begin{align}
&    {\cal E}_0(\Lambda, L) /N
=\left\{\tilde \Lambda^2+ \frac{\partial }{\partial x }\left(
\frac{\partial }{\partial x' }-\frac{\partial }{\partial x }\right)
\right\} D(x,\epsilon;x',0)\Big|_{x=x'=0}    \nn
&=     \left\{\tilde \Lambda^2- 2\frac{\partial^2 }{\partial x^2 }\right\} 
\frac1{2\pi}K_0\big(\tilde \Lambda \sqrt{(x-x')^2+\epsilon^2}\big)
\Big|_{x=x'=0}     \nn
&-\tilde \Lambda^2  \, \frac{A }{2\pi} \left(K_0\big(\tilde \Lambda \sqrt{(x+x' +L)^2+\epsilon^2}\big)
+K_0\big(\tilde \Lambda \sqrt{ (L-x-x')^2+\epsilon^2}\big)\right) 
\Big|_{x=x'=0} +\cdots 
\end{align}
Now
\begin{eqnarray}     -   2\frac{\partial^2 }{\partial x^2 }
\frac1{2\pi}K_0\big(\tilde \Lambda \sqrt{(x-x')^2+\epsilon^2}\big)
\Big|_{x=x'}  =   \frac{ K_1 (\tilde \Lambda \epsilon)}{\pi}\, \frac{\tilde \Lambda}{\epsilon}\;, 
\end{eqnarray}
so that
\begin{eqnarray} 
&&  \left\{{\tilde \Lambda}^2- 2\frac{\partial^2 }{\partial x^2 }\right\} 
\frac1{2\pi}K_0\big(\tilde \Lambda \sqrt{(x-x')^2+\epsilon^2}\big)
\Big|_{x=x'= 0}   =  \frac{{\tilde \Lambda}^2 }{2\pi}  \left[ K_0 (\tilde \Lambda \epsilon) +  2 \frac{ K_1 (\tilde \Lambda \epsilon)}{\tilde \Lambda \epsilon} \right]    \nn
&=&   \frac{{\tilde \Lambda}^2}{ 2  \pi}   K_2 (\tilde \Lambda \epsilon)\;,
\end{eqnarray}
where the identity    
\be   K_2(x) -   K_0(x) =  \frac{2 K_1(x)}{x} \;,
\ee
has been used. Finally
\begin{eqnarray} 
 {\cal E}_0(\Lambda, L) 
&=&  N \left[  \frac{\tilde \Lambda^2}{2\pi} K_2(\tilde \Lambda \epsilon)
-\frac{\tilde \Lambda^2 A}{\pi} K_0\big(\tilde
\Lambda\sqrt{L^2+\epsilon^2}\big)+
{\cal O}\big(e^{-2\tilde \Lambda L}\big)  \right]   \nn
& = &     \frac{N }{\pi \epsilon^2}  -  \frac{ N   {\tilde \Lambda}^2}{4 \pi} -\frac{ N     \tilde \Lambda^2 A}{\pi} K_0(\tilde
\Lambda  L )+
{\cal O}\big(e^{-2\tilde \Lambda L}\big)   +  {\cal O}(\epsilon^2)   \nn
&=&   \frac{N}{\pi\epsilon^2}
-\frac{N \Lambda^2}{4\pi}
-\frac{N A \Lambda^2}{\pi}\left(K_0(\Lambda L) +K_2(\Lambda L)\right)   +{\cal O}\big(e^{-2\tilde \Lambda L}\big)  +  {\cal O}(\epsilon^2) \;,  \label{theresult}
\end{eqnarray}
where    $K_2(z)\sim  \tfrac{2}{z^2} -   \tfrac{1}{2} + \mathcal{O}(z^2)$ at small $z$, and  we  made the replacement 
\be      \tilde \Lambda^2
\equiv  \Lambda^2 e^{2a}  \simeq    \Lambda^2  ( 1+ 2a) \sim    \Lambda^2   (1+4 A K_2(\Lambda L))\;, \label{LamdtildBis}
\ee
in the last line.   In the $L\to \infty$  limit  (\ref{theresult})
approaches  the function $ {\cal E}_{0} (\Lambda, L=\infty) $, calculated in  Appendix \ref{sec:infiniteL},
exponentially fast.

\section{WKB analysis  \label{sec:WKB}}

Assume that for a given value of $L$,  $\lambda(x)$ has been found.  
We adopt the WKB approximation to  the Schr\"odinger equation
\be 
 -f''_n(x)+\lambda(x)f_n(x)=\omega_n^2 f_n(x)\;,\qquad 
\int_{-L/2}^{L/2} dx \; f_n(x)f_m(x)=\delta_{n,m} \;,    \label{usingthis}
\ee
   in order to study the nature of the divergences in
   \be   {\cal E}_0(x)  \equiv  \frac{N}{2}  \sum_{n}\left(\omega_n f_n(x)^2+\frac{1}{\omega_n}f'_n(x)^2  
	 \right)e^{-\epsilon \omega_n}   +\sigma'(x)^2  \;,    \label{EdensityConstBis}
\ee
 i.e., the high-$n$ behavior of the summand.
   As ${\cal E}_0$ is constant in $x$, we shall set  $x= 0$, where  $\sigma'(0)=0$.

The WKB quantization condition is given by\footnote{Due to the sharp rise of the potential $\lambda(x)$ near the boundaries, the phase shift in the WKB wave function is 
$0$ rather than  $\frac{\pi}{4}$  (the Maslov index  being $0$ rather than $1$).
One has   $n$ instead of the familiar  $n+ \frac{1}{2}$ on the right hand side of  (\ref{maslov}). The situation is analogous to the case of the rigid wall. 
We thank G. Paffuti for discussions   on this point.    } 
 \be  2   \int_a^b dx \;  p(x) = 2 \pi  \,  n   \;,  \qquad   p(x)=   \sqrt{\omega_n^2-  \lambda(x)}\;; \qquad   n  \in {\mathbbm Z}_{\ge 0}\;,    \label{maslov}
\ee
\be      p(a)=p(b)=0\;;
\ee
where $a \sim  - \tfrac{L}{2}$,  $b \sim  \tfrac{L}{2}$ for large $n$.    The wave function and its derivative are given by
\be  f_n(x)=   \frac{C}{\sqrt {p(x)}}    \cos \left( \int_a^x  p(x) dx  -  \frac{\pi}{2}   \right)\;.
\ee
\be  f_n^{\prime}(x)=  -  \frac{C   p(x) }{\sqrt {p(x)}}    \sin \left( \int_a^x  p(x) dx  -  \frac{\pi}{2}   \right)   -    \frac{C   p'(x)}{ 2 \big(\sqrt {p(x)}\big)^3}    \cos \left( \int_a^x  p(x) dx  -  \frac{\pi}{2}   \right)   \;.
\ee
\be   \frac{C^2}{2}  \int_a^b  \dfrac{dx}{p(x)}  =1\;,
\ee
to first order in $\hbar$ (implicit here).
Near the boundaries  $\lambda(x)$ behaves as 
\beq
\lambda(x) \simeq    \frac{1}{2  \,(x\pm L/2)^2 \log{1/|x \pm L/2|}}  \;,    \qquad  x\sim \mp  \frac{L}{2} \;,  \label{behavesasBis}
\eeq 

Let us check the large-$L$ limit first. There 
\be \lambda(x) \sim \Lambda^2\;;\qquad  p(x)\sim \sqrt{\omega_n^2 - \Lambda^2} \;;
\ee 
\be    L \, p(x) =  L  \sqrt{\omega_n^2 - \Lambda^2} = \pi    n \;;
\ee
\be \omega_n^2 =  \left( \frac{  \pi  n }{L} \right)^2 +  \Lambda^2\;;\qquad p(x)\sim   \frac{  \pi  n }{L} \;; \label{zeroth}
\ee
\be  f_n(0)  \sim  \sqrt{\frac{2}{L} } \sin  \frac {\pi n}{2}\;.
\ee 
This leads to the calculation  for  ${\cal E}_0(\Lambda,L=\infty)$  described in Appendix \ref{sec:infiniteL}.    
To find the corrections,    write
\be   p(x)=  \sqrt{\omega_n^2 - \Lambda^2}  + \delta p(x)\;, \qquad  \delta p(x) =   \sqrt{\omega_n^2 -\lambda(x)} -    \sqrt{\omega_n^2 - \Lambda^2}  <0\;.\label{deltap}
\ee
The quantization condition is corrected  to
\be    L  \sqrt{\omega_n^2 - \Lambda^2}  +    \Delta_n =   \pi    n \;;\qquad     \Delta_n =   \int  \delta p(x)   dx   <0 \;; \qquad |\Delta_n | \ll  L  \sqrt{\omega_n^2 - \Lambda^2} \;.
\ee
As $|\Delta_n| $  is small compared to $n$,  it may be calculated by inserting the zeroth-order WKB for $\omega_n$ (\ref{zeroth}) in (\ref{deltap}). 
Thus
\be   \omega_n^2 =  \left( \frac{  \pi  n   -  \Delta_n}{L} \right)^2 +  \Lambda^2 \simeq 
 \left( \frac{  \pi  n }{L} \right)^2 +  \Lambda^2   -  \frac{ 2    \pi  n \Delta_n}{L^2}  \;;  \label{omegan}
\ee 
\be  \delta  p(0) \simeq   0\;;
\ee
\be   \int_a^b  \dfrac{dx}{p(x)}   \simeq  \frac{L}{\sqrt{\omega_n^2 -  \Lambda^2}} -  \frac{\Delta_n }{\omega_n^2 - \Lambda^2}\;.
\ee
A straightforward calculation leads to
\be    \omega_n   f_n^2    \simeq    \frac{2}{L}    \frac{  \pi  n  }{L}    \left(  1+     \frac{\Lambda^2 L^2}{2   (\pi n)^2}  \right)  
\left(    1 +   \frac{\Delta_n \Lambda^2 L^2}{  (\pi n)^3}   \right) \sin^2  \left(  \frac{\pi n}{2}  \right) \;;  \nonumber 
\ee
\be \frac{  f_n^{\prime \, 2} }{\omega_n} \simeq  \frac{2}{L}    \frac{  \pi  n  }{L}    \left(  1-     \frac{\Lambda^2 L^2}{2   (\pi n)^2}  \right)  
\left(    1 -   \frac{\Delta_n \Lambda^2 L^2}{  (\pi n)^3}   \right) \cos^2  \left(  \frac{\pi n}{2}  \right) \;,   \label{WKB2}
\ee
in the region
$    \left(  \frac{  \pi  n  }{L}  \right)^2   \gg  \Lambda^2\;
$

The last ingredient needed is  the large $n$ behavior of $\Delta_n$.
It is easy to estimate
 \be \Delta_n  \sim   -  c_1  -  \frac{  c_2  \Lambda^2 L^2}{ 2 \pi n}\;,   \qquad  c_1 \sim \mathcal{O}(1), \quad c_2 \ll1\;,  \label{which11}   \ee  
 at large $n$.
It follows from  (\ref{WKB2}), (\ref{which11}) that  
\be  \omega_n f_n(x)^2+\frac{1}{\omega_n}f'_n(x)^2  
  \sim   C_1  \,  n  +   C_{-2}  \, n^{-2}   +\mathcal{O}(n^{-3})\;,
\ee
at large $n$, where  $C_1$ and $C_{-2}$ are constants of order of unity.
   No $n^0$ and  $n^{-1}$ terms appear.  Thus the divergences in   ${\cal E}_0$ is purely quadratic and is equal to   $ \frac{N}{\pi \epsilon^2}$,  the same as in  the $L\to \infty$   
system.

\section{Random walk algorithm \label{sec:walk}}

In this appendix, we will describe the algorithm we have used for the
numerical calculations in more detail using pseudo-code. 

For the numerical calculations, we can only include a finite number of
modes in the sum in the left-hand equation in Eq.~\eqref{gapeqbb},
henceforth we shall denote this number as {\tt nmax}.
Next we have to discretize all the numerical functions on the interval
on a lattice with {\tt LEN} lattice points, which we for convenience
will take to be an odd integer.
As explained in the text, we will use the symmetry of the problem to
make $\lambda$ manifestly symmetric (with respect to $x\to -x$) in the
calculation.

Take a $\lambda$ which is guessed or just $\lambda=1$ (we started
indeed with this)
\begin{lstlisting}
LENHALF = ceil(LEN/2)
lambda = ones(LEN)
\end{lstlisting}
Now we need a function to calculate the error of using the current
$\lambda$ as compared to the true solution.
We will define the function
\begin{lstlisting}
function err = lambdaerr(lambda)
\end{lstlisting}
First we calculate $\sigma$ from the equation of motion
\eqref{gapeqbb}: 
\begin{lstlisting}
  sigma1 = Delta\[ BC ; zeros(LEN-2) ; BC ]
\end{lstlisting}
where {\tt Delta} is the discretized second-order differential
operator.
The '$\backslash$' notation is an implemented operator in MATLAB and 
Octave for a linear-algebra operation sometimes called back solving.
Formally it is equivalent to multiplying by the matrix inverse of
{\tt Delta} from the left. Numerically, however, that is much more
computationally expensive and hence one should instead use a back
solving algorithm. In Mathematica it is implemented as a function
called {\tt LinearSolve}. 
Then we calculate $\sigma$ again using the gap equation
\begin{lstlisting}
  sigma2sq = r
  [V,D] = eigensystem(Delta)
  for i = 1:nmax
    fn = V(i)/(sqrt(hx*sum(V(i)*V(i))))
    sigma2sq = sigma2sq - fn^2/(2*sqrt(D(i)))
  end for
\end{lstlisting}
Most computational packages have a built-in function for finding the
eigenvectors and eigenvalues of a given matrix, here we will call it
{\tt eigensystem} and denote by {\tt V} the eigenvectors and by
{\tt D} the eigenvalues. Other programming languages have libraries
for linear algebra manipulations that include such a function,
e.g.~{\tt LAPACK} for Fortran90 or {\tt CLAPACK} for C.
Now calculate the error as
\begin{lstlisting}
  err = hx*sum(abs(sigma1^2 - sigma2sq))
end function
\end{lstlisting}
Start the algorithm
\begin{lstlisting}
err = 1
errtol = 1e-5
while (err > errtol) do
\end{lstlisting}
Randomly select an interval that should be changed
\begin{lstlisting}
  istart = round(LENHALF*random())
  istop = round(LENHALF*random())
\end{lstlisting}
Decompose $\lambda$ into a difference vector
\begin{lstlisting}
  diffvec = lambda(LENHALF-1:end-1) - lambda(LENHALF:end)
\end{lstlisting}
Act on the selected range with a random multiplication factor and a
random addition
\begin{lstlisting}
  diffvec = [diffvec(1:istart-1),
    scalefactor*random(istop-istart+1)
      .*(diffvec(istart:istop) + additionfactor*random(istop-istart+1)),
    diffvec(istop+1:end)]
\end{lstlisting}
where {\tt .*} denotes an inner product on the vector space.
The addition factor is necessary in the beginning if one chooses to
start with $\lambda=1$. At the end of the convergence, it should be
small or turned off.\footnote{In order to improve the convergence, we
  have implemented some tweaks for the midpoint.}
Now reconstruct the new $\lambda$ from the difference vector
\begin{lstlisting}
  newlambdahalf = cumsum(diffvec)
  newlambda = [flip(newlambdahalf(2:end)),newlambdahalf]
\end{lstlisting}
where {\tt cumsum} denotes a function that sums cumulatively. 
Test the new $\lambda$:
\begin{lstlisting}
  temperr = lambdaerr(newlambda)
  if (temperr > err) then
    lambda = newlambda
    err = temperr
  end if
end while
\end{lstlisting}
If the discrepancy between the $\sigma$ calculated from the equation
of motion and the $\sigma$ calculated from the gap equation has
decreased, then store the new $\lambda$ and continue; on the other
hand, if the error has increased, then discard the new step and try
again.
The cycle continues until the error is small enough (set by {\tt
  errtol}). 

Various small tweaks can be implemented in the algorithm depending on
the part of parameter space one is interested in. Those tweaks,
however, just make the algorithm converge faster, but to the same
solutions.

We should mention that if one suspects that the guess will converge to
a local vacuum and not to the true vacuum of the functional space,
then the metropolis algorithm can be used to accept increases in the
error at an initial stage of the random walk. When the error decreases 
or when the running time increases, this allowance of ``going in the
wrong direction'' should then be decreased.
On the grounds of knowing the solutions from Ref.~\cite{BKO}, we have
not used this possibility in most of the calculations.

\section{Algorithm test\label{app:algorithm_test}}

In this appendix, we will test the algorithm by choosing a poor
initial condition, i.e.~$\lambda_{\rm ini}(x)=0$, and check which
solution the algorithm will find.
Most solutions presented in the text were found by starting with a
much better guess for $\lambda$. 

Slightly more advanced than what is described in App.~\ref{sec:walk},
we will run the algorithm on a computing cluster and only the best
improvement of each cycle will be accepted.

\begin{figure}[!htp]
\begin{center}
\includegraphics[width=0.32\linewidth]{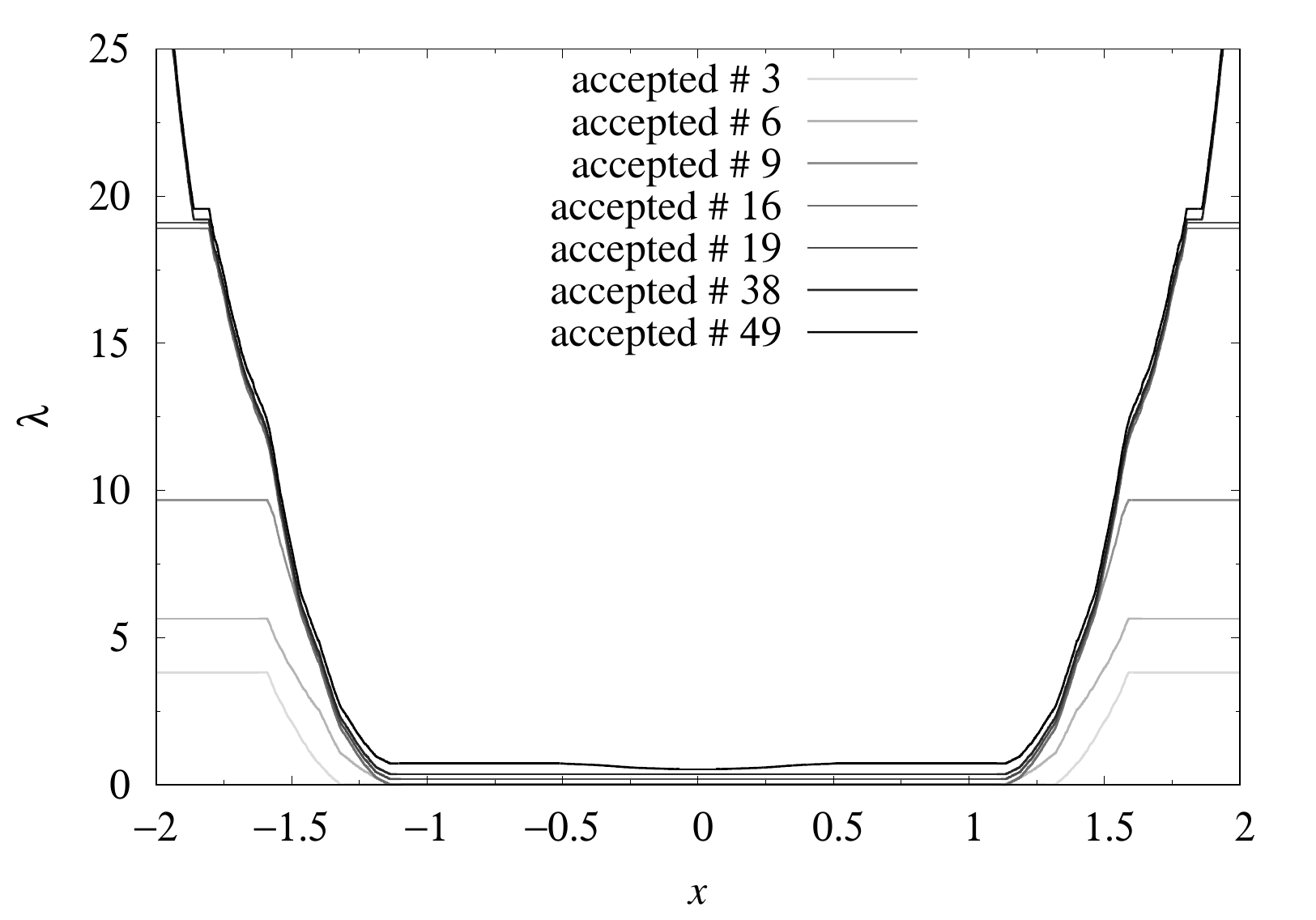}
\includegraphics[width=0.32\linewidth]{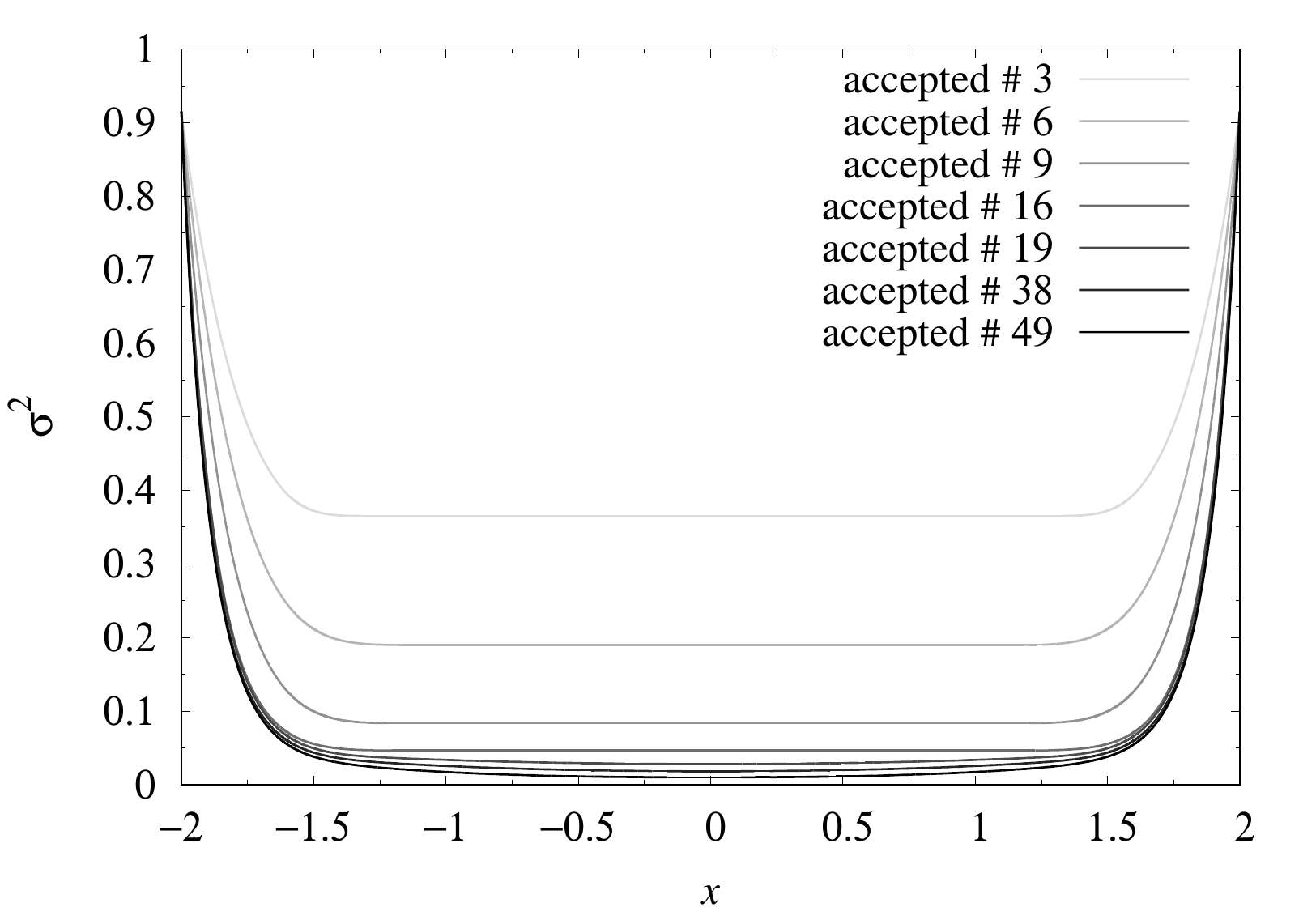}
\includegraphics[width=0.32\linewidth]{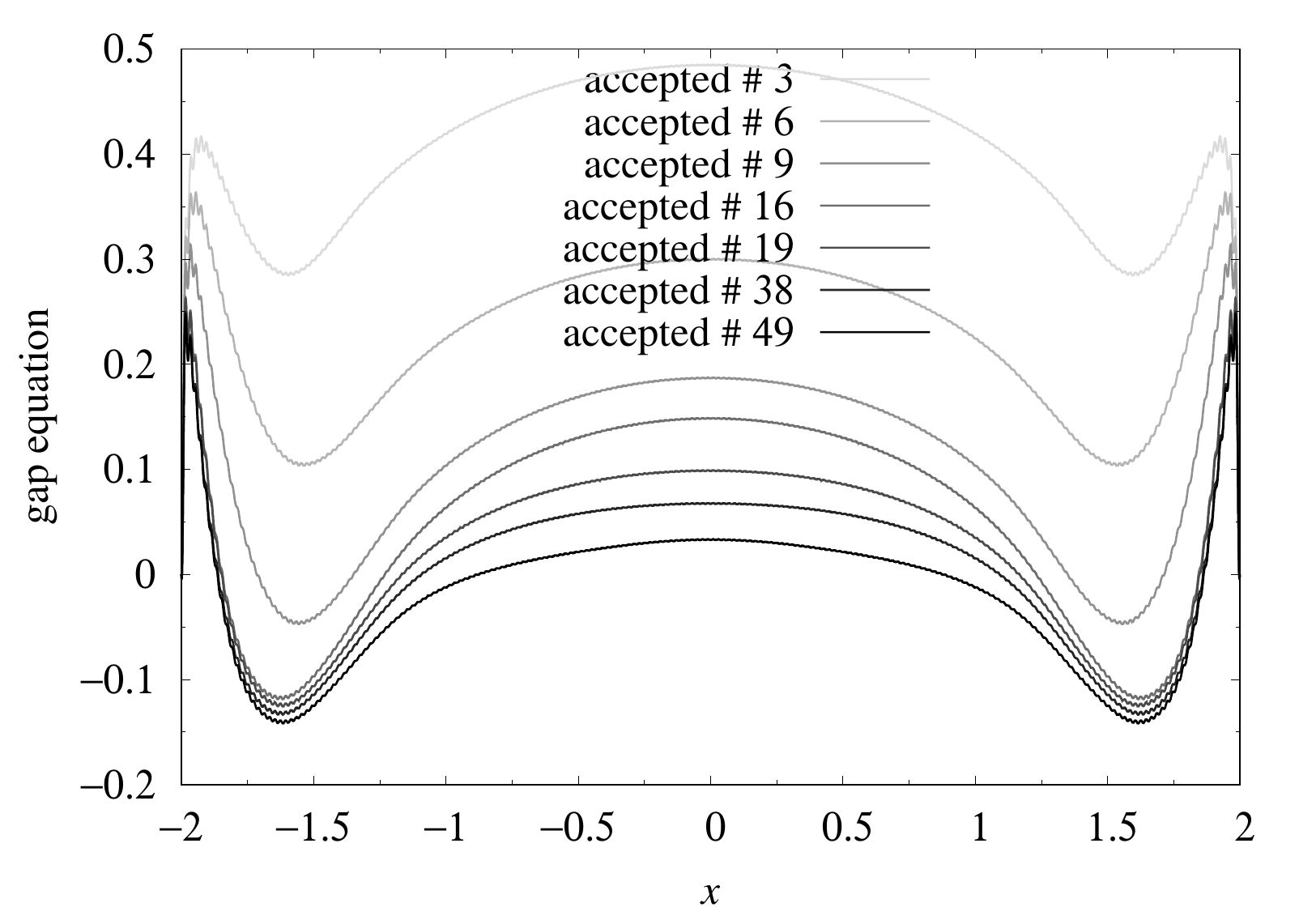}
\caption{Initial stage of the random walk: the algorithm tries to
reduce the numerical error the most by bringing down $\sigma(0)$
towards zero. } 
\label{fig:Higgstest1}
\end{center}
\end{figure}

Since the algorithm prefers the largest decrease in the numerical
error ({\tt err}) at all times, the first thing it wants to do is to
bring down $\sigma(0)$ towards zero.
This happens very quickly by randomly adding arbitrary values to
$\lambda$ near the border, see Fig.~\ref{fig:Higgstest1}.
Recall that the algorithm is programmed to make $\lambda$ a
monotonically increasing function on the interval $[0,L/2]$.
The algorithm randomly chooses where and how much to increase the
function and uses the gap equation to accept or discard the random
steps. 

\begin{figure}[!htp]
\begin{center}
\includegraphics[width=0.32\linewidth]{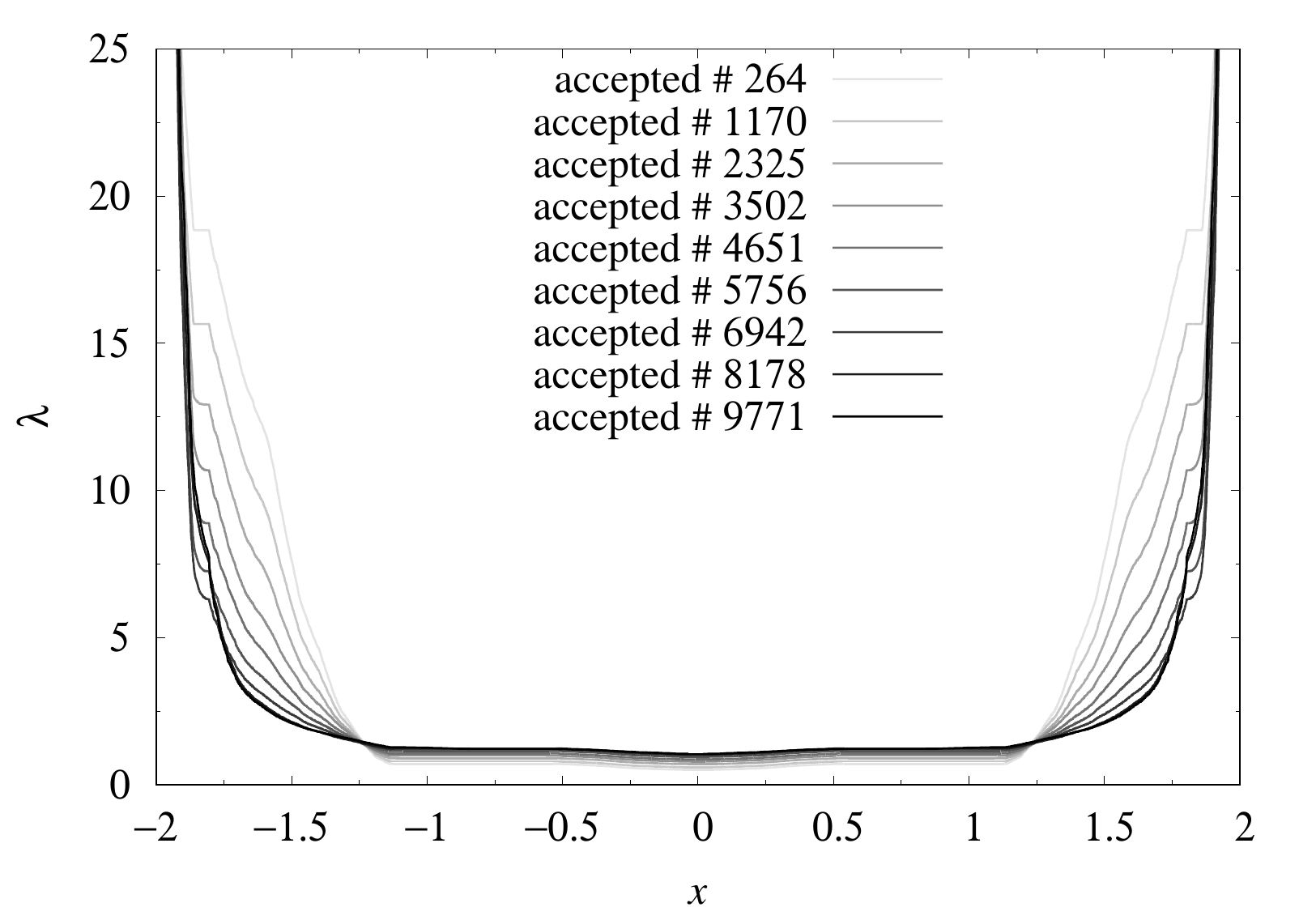}
\includegraphics[width=0.32\linewidth]{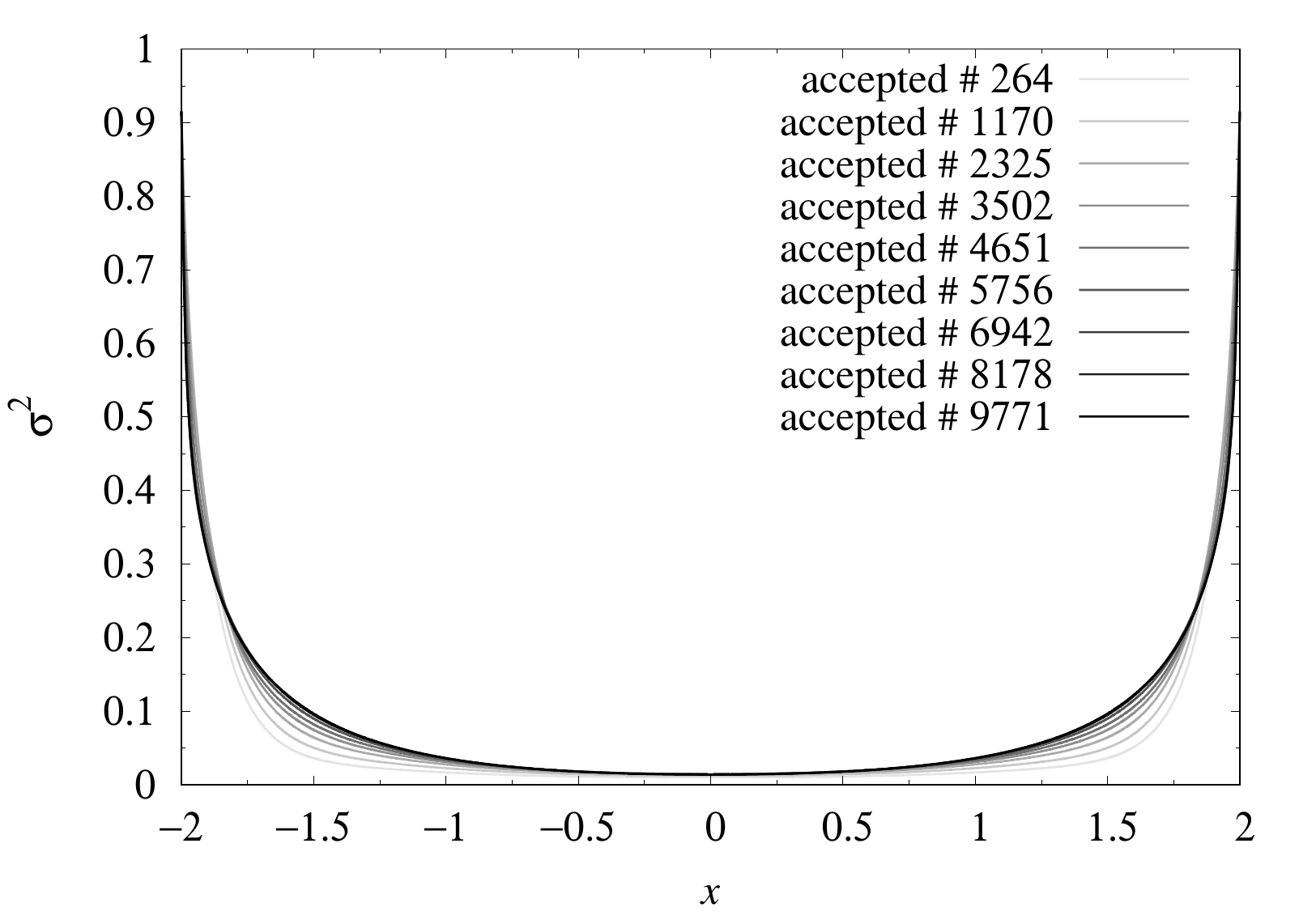}
\includegraphics[width=0.32\linewidth]{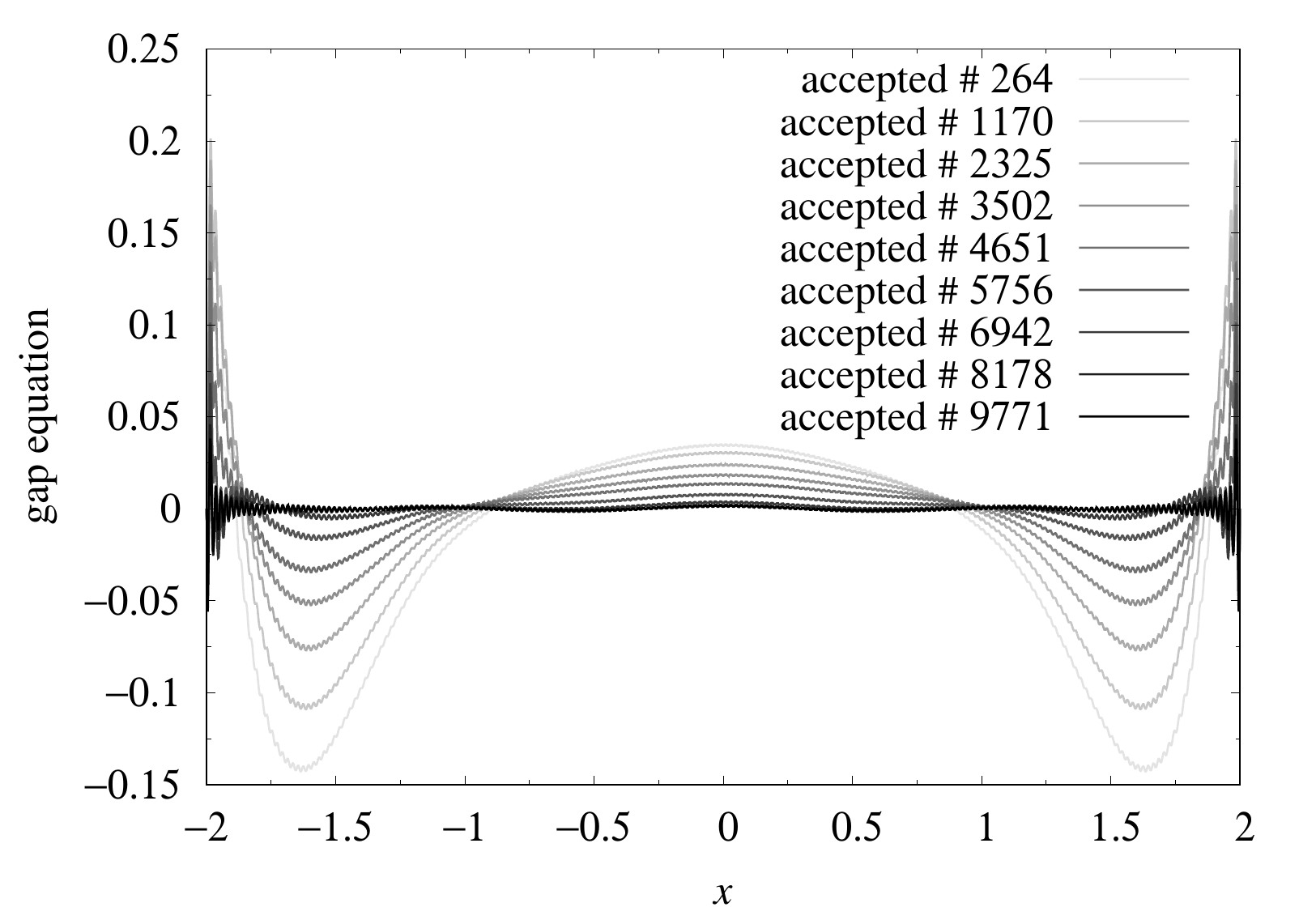}
\caption{Later stage of the random walk: the algorithm reduces the
numerical error by ``pushing out'' the corners of $\lambda$ and
adjusting $\lambda(0)$. }
\label{fig:Higgstest2}
\end{center}
\end{figure}

Unfortunately, the randomly chosen (by the algorithm) values of
$\lambda$ near the boundary yield too ``sharp'' a solution for
$\sigma$; to mitigate this, the algorithm sees the numerical error can
be reduced by ``pushing out'' the corners of $\lambda$ and adjusting
the midpoint, $\lambda(0)$, which after enough cycles yields a
solution for $\sigma$ to the gap equation and hence a solution for
$\lambda$, see Fig.~\ref{fig:Higgstest2}. 
The algorithm terminates when the error is below a given acceptable
threshold ({\tt errtol}).
The solution is shown as a black line in Fig.~\ref{fig:Higgstest2};
i.e.~this solution has been accepted with an error tolerance of
{\tt errtol} $=6\times 10^{-5}$.

\begin{figure}[!htp]
\begin{center}
\includegraphics[width=0.49\linewidth]{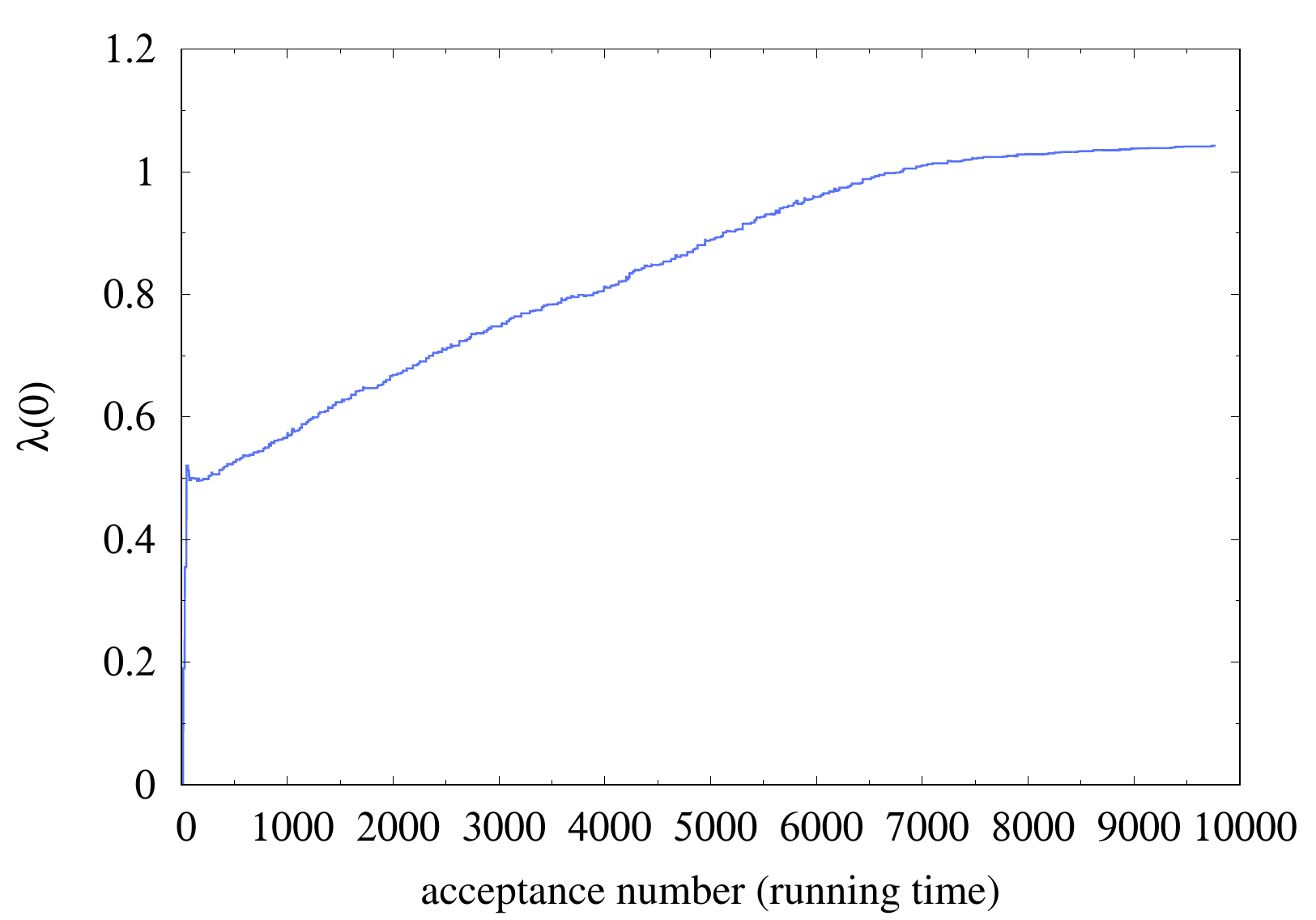}
\includegraphics[width=0.49\linewidth]{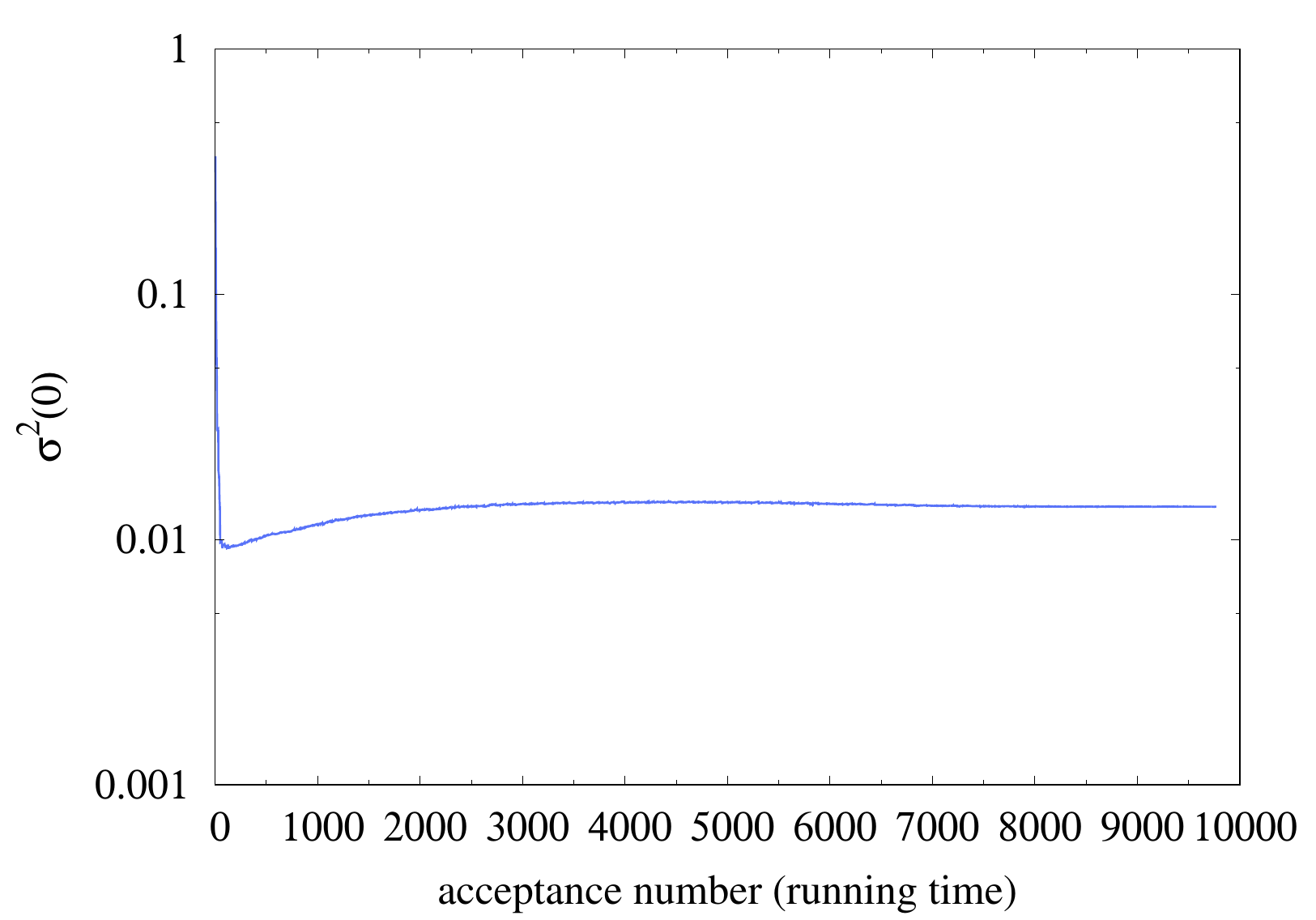}
\caption{The value of $\lambda$ and $\sigma$ at the midpoint of the
string as functions of acceptance number (running time). }
\label{fig:Higgstest_midpoints}
\end{center}
\end{figure}

Finally, in Fig.~\ref{fig:Higgstest_midpoints} we display the midpoint
values of $\lambda$ and $\sigma^2$ as functions of acceptance numbers
(which roughly corresponds to running time of the numerical
calculation). 

\end{document}